\documentclass[structabstract]{aa}

\usepackage[colorlinks=true,linkcolor=black, linktocpage, citecolor  = black, urlcolor = black, 
bookmarksopen]{hyperref}
\usepackage{amsmath}
\usepackage{natbib}
\usepackage{url}
\usepackage{units}
\usepackage{color}

\usepackage{graphicx}
\usepackage{txfonts}

\usepackage{subfigure}

\usepackage{framed}

\newcommand{\pd}[2]{\frac{\partial #1}{\partial #2} }
\newcommand{\pdl}[2]{\frac{\partial}{\partial #2} #1 }
\newcommand{\pdld}[2]{\frac{\partial^2}{\partial #2^2} #1 }

\newcommand{\ind}{ {i,j,k} }
\newcommand{\indv}[3]{ {i#1,j#2,k#3} }
\newcommand{\gf}[2]{G_{#1}^{#2}}
\newcommand{\oh}{\frac{1}{2}}

\newcommand{\id}{\,\mathrm{d}}

\newcommand{\acc}{\vec{a}}
\newcommand{\accRad}{\acc_\mathrm{rad}}
\newcommand{\accExt}{\acc_\mathrm{ext}}

\newcommand{\moH}{m_\mathrm{H}}

\newcommand{\PLUTO}{{\tt PLUTO}\,}

\newlength{\ibwidth}
\newlength{\iswidth}
\begin{document}
	\title{Radiation hydrodynamics integrated in the code PLUTO}
	\author{Stefan M. Kolb
 		\inst{1}
	\and
		Matthias Stute
 		\inst{1}
	\and
		Wilhelm Kley
 		\inst{1}
	\and
		Andrea Mignone
 		\inst{2}
	}

	\institute{
	Institute for Astronomy and Astrophysics, Section Computational Physics, Eberhard Karls 
Universit\"at 
	T\"ubingen, Auf der Morgenstelle 10, D-72076 T\"ubingen, Germany\\
	\email{kolb.stefan@gmail.com, matthias.stute@uni-tuebingen.de, 
wilhelm.kley@uni-tuebingen.de}\\
	\and
        Dipartimento di Fisica, Universit\`a degli Studi di Torino, via Pietro Giuria 1, 10125 
Torino, Italy
	}

 	\date{Received ; accepted 9. September 2013}

\abstract
        {}
        {The transport of energy through radiation is very important in many astrophysical phenomena.
        In dynamical problems the time-dependent equations of radiation hydrodynamics have to be solved.
 	We present a newly developed radiation-hydrodynamics module specifically designed for the versatile MHD code \PLUTO.}
 	{The solver is based on the flux-limited diffusion approximation in the two-temperature approach.
        All equations are solved in the co-moving frame in the frequency independent (grey) approximation.
        The hydrodynamics is solved by the different Godunov schemes implemented in \PLUTO, and for the radiation transport we use a fully
        implicit scheme. The resulting system of linear equations is solved either using the successive over-relaxation (SOR) method
        (for testing purposes), or matrix solvers that are available in the PETSc library.
	We state in detail the methodology and describe several test cases in order to verify the 
	correctness of our implementation.
        The solver works in standard coordinate systems,
        such as Cartesian, cylindrical and spherical, and also for non-equidistant grids.
        }
	{We have presented a new radiation-hydrodynamics solver coupled to the MHD-code \PLUTO that is a modern, versatile and efficient 
 	new module for treating complex radiation hydrodynamical problems in astrophysics. 
        As test cases, either purely radiative situations, or full radiation-hydrodynamical setups (including radiative shocks and convection
        in accretion discs) have been studied successfully. The new module scales very well on parallel computers using MPI.
        For problems in star or planet formation, we have added the possibility of irradiation by a central source.
        }
          {}
   \maketitle

\section{Introduction}

Radiative effects play a very important role in nearly all astrophysical fluid flows, ranging from
planet and star formation to the largest structures in the universe.
Coupling the equations of radiation transport to those of (magneto-)hydrodynamics (MHD) has been studied for decades, and 
comprehensive treatments can be found for example in text books by \cite{1984oup..book.....M} or \citet{1973erh..book.....P}. 
The numerical implementation of two-temperature radiation hydrodynamics (in the diffusion approximation)
into multi-dimensional MHD/HD-codes has been done already over twenty years ago in various implementations,
for example by \citet{1988ApJ...330..142E}, \citet{1989A&A...208...98K}, in the {\tt ZEUS}-code \citep{1992ApJS...80..819S},
and later by \citet{2001ApJS..135...95T}.

In order to study, for example, the dynamics and characteristics of stellar atmospheres together with convection,
more accurate solvers for the radiation transport based on the method of short characteristics have been developed,
see \citet{2012ApJS..199....9D} and \citet{2012JCoPh.231..919F} for the present status. This can then be coupled to the hydrodynamics using the
{\it Variable Eddington Tensor} method \citep{2012ApJS..199...14J}. 
Another approach is the $\mathrm{M}1$ closure model where the radiative moment equations are closed
at a higher level \citep{2007A&A...464..429G,2008MNRAS.387..295A}. 
Despite this progress it is still useful and desirable
to have a method at hand which solves the interaction of matter and radiation primarily within the bulk part of the
matter which may be optically thick. In such type of applications, the method of flux-limited diffusion (FLD, see \citet{1981ApJ...248..321L}) has its clear merits and
is still implemented into existing MHD-codes, for example in {\tt NIRVANA} \citep{2009A&A...506..971K} to study the planet formation
process, in {\tt RAMSES} \citep{2011A&A...529A..35C} for protostellar collapse simulations,
and in combination with a multi-frequency irradiation tool into {\tt PLUTO} \citep{2010A&A...511A..81K} for massive star formation.

Since the 3D-MHD code {\tt PLUTO} \citep{2007ApJS..170..228M} is becoming increasingly popular within the computational astrophysics
community, we added a publicly available radiation module, which is based on the two-temperature FLD-approximation,
as described by \citet{2011A&A...529A..35C}. \PLUTO solves the equations of hydrodynamics and magnetohydrodynamics including the non-ideal effects of viscosity, thermal conduction and resistivity by means of shock-capturing Godunov-type methods. Several Riemann solvers, several time stepping methods and interpolation schemes can be chosen.
Additionally, we added a ray-tracing routine that allows for additional irradiation 
by a point source in the center. Treating the irradiation in a ray-tracing approach, guarantees the long-range character
of the radiation better than FLD \citep{2012A&A...537A.122K,2013A&A...555A...7K}.

The paper is organized as follows. In section \ref{sec:equations}, we briefly introduce the equations of hydrodynamics including radiation transport.
Additionally we describe the general idea behind the flux limited diffusion approximation. In section \ref{sec:solving_the_radiation_part}, we present 
the discretization of the equations and the solver of the resulting matrix equation, and present our numerical
implementation of irradiation. In section \ref{sec:test_cases}, we present six different test cases to show the correctness 
of the implemented equations: four test cases with an analytical solution (section \ref{sec:linear_diffusion_test} to
\ref{sec:steady_state_test}) and two others in which our results are compared with those from other codes (section \ref{sec:radiation_shock} and \ref{sec:accretion_disc}).
We end with a summary and conclusions.

\section{Radiation hydrodynamics} 
\subsection{The equations} \label{sec:equations}

Even though the {\tt PLUTO}-environment includes the full MHD-equations and non-ideal effects such as viscosity,
we restrict ourselves here to the Euler equations of ideal hydrodynamics. 
Radiation effects are included in the two-temperature approximation,
which implies an additional equation for the radiation energy.
In order to follow the transport of radiation, we apply the flux-limited diffusion approximation
and treat the exchange of energy and momentum between the gas and the radiation field with additional terms
in the gas momentum and energy equations. The system of equations then read:

\begin{flalign}
\pdl{\rho}{t} + \nabla  \cdot (\rho \vec{v}) &= 0 \label{eqn:continuity_equation} \\
\pdl{\rho \vec{v}}{t}+\nabla \cdot \left(\rho \vec{v} \otimes \vec{v}\right) + \nabla p &=  + \rho 
(\accExt + \accRad)  \label{eqn:momentum_equation_coupled}\\
\pdl{e}{t}+\nabla\cdot\left[\left(e + p\right)\vec{v}\right] & =  + \rho \vec{v} \cdot (\accExt + \accRad)     
\label{eqn:gas_energy_equation_coupled} \\ 
  &\phantom{=}\hspace{3pt} -\kappa_\mathrm{P} \rho c ( a_\mathrm{R} T^4 - E) \nonumber \\
\pdl{E}{t} + \nabla \cdot \vec{F}  &= \kappa_\mathrm{P}\rho c \left( a_\mathrm{R} T^4 - E\right)
\label{eqn:radiation_energy_equation_coupled}
\end{flalign}
The first three equations (\ref{eqn:continuity_equation}-\ref{eqn:gas_energy_equation_coupled})
describe the evolution of the gas motion, where $\rho$ is the gas density, 
$p$ the thermal pressure, $\vec{v}$ the velocity, $e = \rho \,\epsilon + 1/2 \,\rho\,v^2$
the total energy density (i.e., the sum of internal and kinetic) of the gas without radiation, and 
$\accExt$ an acceleration caused by external forces (e.g. gravity), not induced by the radiation field (see below).
This system of equations is closed by the ideal gas relation
\begin{equation}
   p=( \gamma - 1 ) \,\rho \,\epsilon = \rho \frac{k_\mathrm{B}\,T}{\mu\, \moH}, \label{eqn:convert_temperature_to_pressure}
\end{equation}
where $\gamma$ is the ratio of specific heats, $T$ the gas 
temperature, $k_\mathrm{B}$ the Boltzmann constant, $\mu$ the mean molecular weight, and $\moH$ the mass of 
hydrogen. The specific internal energy can be written as $\epsilon = c_\mathrm{V}\,T$, with the specific heat 
capacity given by 
\begin{equation}
\label{eq:cv}
c_\mathrm{V}=\frac{k_\mathrm{B}}{(\gamma - 1) \mu \moH} .
\end{equation}
Here, we assume constant $\gamma$ and $\mu$, which
also implies a constant $c_\mathrm{V}$.

The evolution of the radiation energy density $E$ is given by eq.~\eqref{eqn:radiation_energy_equation_coupled},
where $\vec{F}$ denotes the radiative flux, $\kappa_\mathrm{P}$ the Planck 
mean opacity, $c$ the speed of light and $a_\mathrm{R}$ the radiation constant.
The fluid is influenced by the radiation in two different ways. 
First, the radiation may be absorbed or emitted by the fluid leading to variation of its energy density.
This variation is given by the expression $\kappa_\mathrm{P}\rho c \left( a_\mathrm{R} T^4 - E\right)$,
see right hand side of equation \eqref{eqn:gas_energy_equation_coupled} and \eqref{eqn:radiation_energy_equation_coupled}.
The second effect is that of radiation pressure. We 
include this term as an additional acceleration to the momentum equation, $\accRad = 
\frac{\kappa_\mathrm{R}}{c} \vec{F}$.
The present implementation does not include the advective transport terms for the radiation energy and  radiative pressure work in 
eqs.~\eqref{eqn:gas_energy_equation_coupled} and \eqref{eqn:radiation_energy_equation_coupled}. 
For the relatively low temperature protoplanetary disk application
that we consider here these terms are of minor importance.
If required, these terms can be treated in our implementation straightforwardly within {\tt PLUTO} by adding additional source terms.

\subsection{The flux-limited diffusion approximation}

The system of equations shown cannot be solved without further assumptions for
the radiative flux $\vec{F}$. Here we use the flux limited diffusion approximation (FLD) where
the radiation flux is given by a diffusion approximation
\begin{equation}
\vec{F} = - \lambda \, \frac{c}{\kappa_\mathrm{R}\, \rho}\, \nabla E \label{eqn:radiative_flux} \,,
\end{equation}
with the Rosseland mean opacity $\kappa_{\rm R}$. The flux-limiter $\lambda$
describes approximately the transition 
from very optically thick regions with $\lambda = 1/3$ to optically thin regimes, where
$\vec{F} \rightarrow - c E \frac{\nabla E}{|\nabla E|}$. This leads to the formal definition of the 
flux-limiter which is a function of the dimensionless quantity 
\begin{equation}
  R=\frac{|\nabla E|}{\kappa_\mathrm{R} \rho E} \,,   \label{eqn:R}
\end{equation}
with the following behaviour:
\begin{equation}
\lambda(R) = \left\{\begin{array}{ll}\frac{1}{3}, & R \to 0 \\ 
\frac{1}{R}, & R \to \infty \end{array}\right. \label{eqn:fluxlimiter_limits}
\end{equation}
Physically sensible flux-limiters thus have to fulfil the equation \eqref{eqn:fluxlimiter_limits} 
in the given limits and describe the behaviour between the limits approximately. We have implemented three different 
flux-limiters:
\begin{flalign}
\lambda(R) &= \frac{1}{R}\left(\coth R - \frac{1}{R}\right) \label{eqn:fluxlimiter_levermore_pomraning}\\
\lambda(R) &= \left\{ 
	\begin{array}{l l}
		\frac{2}{3 + \sqrt{9+12 R^2}} & \quad 0 \leq R \leq \frac{3}{2}\\
		\frac{1}{1+R+\sqrt{1+2R}} & \quad \frac{3}{2} < R \leq \infty\\
	\end{array} \right.\label{eqn:fluxlimiter_minerbo}\\
\lambda(R) &= \left\{ 
	\begin{array}{l l}
		\frac{2}{3 + \sqrt{9+10 R^2}} & \quad 0 \leq R \leq 2\\
		\frac{10}{10 R+ 9 + \sqrt{180 R + 81}} & \quad 2 < R \leq \infty\\
	\end{array} \right.\label{eqn:fluxlimiter_kley}
\end{flalign}
from \cite{1981ApJ...248..321L}, \cite{1978JQSRT..20..541M}, and \cite{1989A&A...208...98K}, respectively.  
A comparison of them is presented in \cite{1989A&A...208...98K}.

In general it is necessary to solve the equations for each frequency which appears in the physical problem. 
However, here we use the grey approximation in which all radiative quantities including the opacities are
integrated over all frequencies. 
In our treatment
scattering is not accounted for directly, but it is included in the effective isotropic absorption and 
emission coefficients.

\section{Solving the radiation part} \label{sec:solving_the_radiation_part}
\subsection{Reformulation of the equations}
Instead of solving system of equations (\ref{eqn:continuity_equation}-\ref{eqn:radiation_energy_equation_coupled})
directly as a whole, the problem is split into two steps.
In the first step, \PLUTO is used to solve the equations of fluid dynamics with the 
additional force caused by the radiation. This corresponds to the equations 
\eqref{eqn:continuity_equation} to \eqref{eqn:gas_energy_equation_coupled} with the additional acceleration,
$\accRad$, but without the interaction term between the matter and radiation (last term in eq.~\ref{eqn:gas_energy_equation_coupled}).
By using \PLUTO for solving the non-radiative part of the equations, we are not limited 
to the Euler equations, but are able to use the full capabilities of \PLUTO for solving the 
equations of hydrodynamics or magnetohydrodynamics, including the effects of viscosity and 
magnetic resistivity. 

In a second, additional step we solve the radiation energy equation (\ref{eqn:radiation_energy_equation_coupled})
and for the corresponding heating-cooling term in the internal energy of the fluid:
\begin{equation}
\left.
\begin{array}{lll}
  \displaystyle {\pdl{E}{t} - \nabla \cdot \left( \frac{c \lambda}{\kappa_\mathrm{R} \rho} \nabla E \right)} &=& \phantom{-} \kappa_\mathrm{P}\rho c \left( a_\mathrm{R} T^4 - E \right) \\ 
 \displaystyle {\pdl{\rho \epsilon}{t}} & = & -\kappa_\mathrm{P} \rho c \left(  a_\mathrm{R} T^4 - E \right)  
\end{array}
\quad \right\}
\label{eqn:energy_coupling}
\end{equation}
In order to obtain the radiation energy density, we solve the system of coupled equations 
\eqref{eqn:energy_coupling}.
Within one time step \PLUTO advances the hydrodynamical quantities, i.e. the density $\rho$, the velocity $\vec{v}$ and a 
temporary pressure $p$ from time $t^{n}$ to the time $t^{n+1}$, where the time step, $\Delta t = t^{n+1} - t^n$, is determined 
by \PLUTO using the CFL conditions, presently without including the radiation pressure. These depend on the used time stepping method in \PLUTO, for more information see
\cite{2007ApJS..170..228M} and the userguide of \PLUTO.

The physical process of radiation transport takes place on time scales much shorter than the one in hydrodynamics.
In order to use the same time step for hydrodynamics and the radiation transport, we apply an implicit scheme to handle the 
radiation diffusion and the coupling between matter and radiation described by equation \eqref{eqn:energy_coupling}.
Because of the coupling of the equations, the method will update $T$ and $E$ simultaneously, which leads formally
to a nonlinear set of coupled equations.
As outlined below, the system is solved for the radiation
energy density $E$.
From the new values for $E$, we compute the new fluid temperature (see eq.~\ref{eqn:calculation_temperature} below) and update the fluid 
pressure by using the ideal gas relation from equation \eqref{eqn:convert_temperature_to_pressure}.
This is then used within \PLUTO to calculate a new total gas energy $e$.

\subsection{Discretization}

In order to discretize the equations \eqref{eqn:energy_coupling}, we 
apply a finite volume method. For that purpose we integrate
over the volume of a grid cell and transform the divergence into a surface integral.
Furthermore, we replace the gradient of $E$ by finite differences, and apply an implicit scheme.
The discretization scheme has been implemented in 3D for Cartesian, cylindrical and spherical polar coordinates
including all the necessary geometry terms for the divergence and gradient.
Since the density has been updated already in the hydrodynamical part of the solver,
we can replace $\pd{\rho \epsilon}{t}$ with $\rho\,c_\mathrm{V}\, \pd{T}{t}$, which is valid for a constant heat capacity.
Then the resulting discretized equations for the radiative part can be written as
\begin{eqnarray}
       &   &  \frac{E_\ind^{n+1} - E_\ind^n}{\Delta t}  \nonumber  \\ 
	&=& \gf{x1}{r} K_\indv{+\oh}{}{}^n \frac{E_\indv{+1}{}{}^{n+1} - E_\ind^{n+1}}{\Delta 
{x_1}_{i+\oh}}
		- \gf{x1}{l} K_\indv{-\oh}{}{}^n \frac{E_\ind^{n+1} - E_\indv{-1}{}{}^{n+1}}{ 
\Delta {x_1}_{i-\oh}} \nonumber \\
	&+& \gf{x2}{r} K_\indv{}{+\oh}{}^n \frac{E_\indv{}{+1}{}^{n+1} - E_\ind^{n+1}}{\Delta 
{x_2}_{j+\oh}}
		- \gf{x2}{l} K_\indv{}{-\oh}{}^n \frac{E_\ind^{n+1} - E_\indv{}{-1}{}^{n+1}}{\Delta 
{x_2}_{j-\oh}} \nonumber \\
	&+& \gf{x3}{r} K_\indv{}{}{+\oh}^n \frac{E_\indv{}{}{+1}^{n+1} - E_\ind^{n+1}}{\Delta 
{x_3}_{k+\oh}}
		- \gf{x3}{l} K_\indv{}{}{-\oh}^n \frac{E_\ind^{n+1} - E_\indv{}{}{-1}^{n+1}}{\Delta 
{x_3}_{k-\oh}}  \nonumber \\
	&+& {\kappa_\mathrm{P}^n}_\ind \, \rho_\ind^n c \left(a_\mathrm{R} (T_\ind^{n+1})^4 - E_\ind^{n+1}\right) \,,
   \label{eqn:discretisized_radiation_energy_equation}
\end{eqnarray}
and for the thermal energy (or temperature, respectively)
\begin{equation}
\frac{T_\ind^{n+1} - T_\ind^n}{\Delta t} = - \frac{{\kappa_\mathrm{P}}_\ind^n \, c}{c_\mathrm{V}} \,  \left(a_\mathrm{R} 
\left(T_\ind^{n+1}\right)^4 - E_\ind^{n+1}\right)\label{eqn:discretisized_fluid_internal_energy_coupling}  \,.
\end{equation}

Here, the superscript $n$ refers to the values of all variables after the most recent update from the hydrodynamical step.
In order to simplify the notation for the separate radiation module, we assume the update takes place from time $n$ to $n+1$.
The subscripts $i,j,k$ refer to the 3 spatial directions of the computational grid, where
all variables are located at the cell centers. Half-integer indices refer to cell interfaces. 
The physical sizes (proper length) of each cell in the 3 spatial directions $m$ ($m=1,2,3$)
are given by $\Delta {x_m}$, where we additionally allow for non-equidistant grids.
The effective radiative diffusion coefficient (defined at cell centers) is given by
\[
 K_\ind^{n} = \frac{c \lambda(R_\ind)}{{\kappa_\mathrm{R}}_\ind^{n} \, \rho_\ind^n} \,,
\]
where $R_\ind$ is calculated from eq.~(\ref{eqn:R}) by central differencing.
Values at cell interfaces are obtained by linear interpolation.
The factors $G^{l,r}_{xm}$ are geometrical terms defined, respectively, as the left and right surface areas
divided by the cell volume in the direction given by $m=1,2,3$.
In the recent work by \citet{2013A&A...549A.124B} the difference equations have been written
out in more detail for Cartesian, equidistant grids. 
The required opacities are evaluated using the values of $\rho$ and $T$ after the hydrodynamical update at time $t^n$.

As mentioned before, equations \eqref{eqn:energy_coupling} constitute a set of coupled nonlinear equations.
The non-linear term $(T_\ind^{n+1})^4$ that appears in equation 
\eqref{eqn:discretisized_fluid_internal_energy_coupling} is linearised using the method outlined in
\cite{2011A&A...529A..35C}
\begin{equation}
(T_\ind^{n+1})^4 = (T_\ind^n)^4\left(1 + \frac{T_\ind^{n+1} - T_\ind^n}{T_\ind^n} \right)^4\approx 
4 (T_\ind^n)^3 T_\ind^{n+1} - 3 (T_\ind^n)^4\,.
\end{equation}
Using this approximation, we obtain an equation for computing the new temperature in terms of the new radiation
energy density, $E_\ind^{n+1}$, and the old temperatures, $T_\ind^n$
\begin{equation}
T_\ind^{n+1} = \frac{{\kappa_\mathrm{P}}_\ind^n c \left(3 a_\mathrm{R} (T_\ind^n)^4  + E_\ind^{n+1}\right)\Delta t + c_\mathrm{V} 
  T_\ind^n}{c_\mathrm{V} + 4 {\kappa_\mathrm{P}}_\ind^n c a_\mathrm{R} (T_\ind^n)^3 \Delta t} \,. \label{eqn:calculation_temperature}
\end{equation}
The expression can be substituted into eq.~(\ref{eqn:discretisized_radiation_energy_equation})
to obtain a linear system of equations for the new radiation energies $E_\ind^{n+1}$, that can be solved
using standard matrix solvers, see section \ref{subsec:matrix-solver}.
The new temperature can then be calculated from eq. \eqref{eqn:calculation_temperature}.
We implemented several boundary conditions for the radiation energy density including periodic, symmetric and fixed value.

\subsection{Irradiation} \label{sec:irradiation}

In order to couple possible irradiation to the radiation transport equations, a new source term, $S$, has to be added 
to the right hand side of the thermal energy equation in system \eqref{eqn:energy_coupling}
\begin{equation}
\pd{\rho \epsilon}{t} = -\kappa_\mathrm{P} \rho c ( a_\mathrm{R} T^4 - E) + S  \,.
   \label{eqn:fluid_internal_energy_coupling_irradiation}
\end{equation}
This results in an additional term, $S_\ind/(\rho_\ind c_{\rm V})$, in Eq.~\ref{eqn:discretisized_fluid_internal_energy_coupling},
correspondingly in equation \eqref{eqn:calculation_temperature}, and in a modification of the right hand side of the resulting matrix 
equation for $E_\ind^{n+1}$.

For the present implementation, we assume that the irradiating source is located at the centre of a spherical 
coordinate system. Therefore it is straightforward to compute the optical depth $\tau_\ind$ even for
simulations using parallel computers. Assuming that a ray of light travels along the radial direction from
the origin to the grid cell $i,j,k$ under consideration, the optical depth from the inner radius $r_0$ to
the $i$th grid cell with radius $r_i$ can be simply expressed as the integral along the radial coordinate,
\begin{equation}
\tau_{i,j,k} = \int_{r_0}^{r_i} \kappa_\star \rho(r) \,dr \approx \sum_{n=0}^{i}  \kappa_\star{_{n,j,k}} \, 
\rho_{n,j,k} \Delta r_n \label{equation::optical_depth}
\end{equation}
where $\Delta r_n$ is the radial length of the $n$th grid cell,
and $\kappa_\star$ the opacity used for irradiation. For the sake of readability, we write $\tau_i$ instead of $\tau_{i,j,k}$ in the following. 
We use $\kappa_\star = \kappa_{\rm P}$ in the test case with irradiation presented in section \ref{sec:coupling_test_irradiation}.
Additionally $\kappa_\star$ can be defined by the user as well as the other opacities.
Re-emission of the photons which were
absorbed in the cell volume is handled in our treatment by the heating-cooling term see equation \eqref{eqn:energy_coupling}.

The luminosity of the source is given by
\begin{equation}
L_\star=4 \pi R_\star^2 \sigma T_\star^4 \,,
\end{equation}
where $\sigma$ denotes the Stefan-Boltzmann constant,
$T_\star$ is the temperature of the star and $R_\star$ its radius.
In order to compute the amount of irradiated energy which is absorbed by a 
specific grid cell we have to know the surface area $A$ of a grid cell oriented perpendicular to 
the radiation from the star and the flux $f$ at the radius $r$. This surface area $A$ is given by 
the expression
\begin{equation}
A_\ind = \int\limits_{\theta_j}^{\theta_{j+1}}\int\limits_{\phi_k}^{\phi_{k+1}} \,d A
= r_i^2 (\phi_{k+1} - \phi_k) (\cos\theta_j -\cos\theta_{j+1}) \,,
\end{equation}
where $\theta$ is the azimuthal and $\phi$ the polar angle in the spherical coordinate system. Without absorption the 
flux $f$ is given by the expression
\begin{equation}
  f = \frac{L_\star}{4 \pi r^2} = \sigma T_\star^4 \left(\frac{R_\star}{r}\right)^2 \,.
\end{equation}
The amount of energy per time which arrives at the surface of the grid cell $(i,j,k)$ is 
\begin{equation}
H_\ind = A_\ind f = (\phi_{k+1} - \phi_k) (\cos\theta_j -\cos\theta_{j+1})  \sigma T_\star^4 R_\star^2\,,
\end{equation}
again without absorption. If the irradiated energy is partly absorbed, the remaining amount of energy per 
time is then $H_\ind e^{-\tau_{i,j,k}}$. Using these results, we can compute the energy density per 
time, $S$, which is absorbed by one grid cell ($i,j,k$)
\begin{flalign}
S_\ind &= \frac{H_\ind e^{-\tau_i} - H_\ind e^{-\tau_{i+1}}}{V_\ind} 
=  \frac{H_\ind \left(e^{-\tau_i}-e^{-\tau_{i+1}}\right)}{V_\ind}\nonumber\\
&=  \frac{3\sigma T_\star^4 R^2_\star \left(e^{-\tau_i}-e^{-\tau_{i+1}}\right)}{(r_{i+1}^3 - 
r_i^3)} \,, \label{eqn:S}
\end{flalign}
with the volume of a grid cell 
\begin{flalign}
V_\ind &= \int\limits_{r_i}^{r_{i+1}} 
\int\limits_{\theta_j}^{\theta_{j+1}}\int\limits_{\phi_k}^{\phi_{k+1}} r^2 sin \theta \, d r \, 
d\theta \, d\phi\nonumber\\
&= \frac{1}{3} (r_{i+1}^3 - r_i^3) (\cos \theta_j - \cos \theta_{j+1}) (\phi_{k+1} - \phi_k)  \,.
\end{flalign}
The absorbed energy density per time, $S_\ind$, is computed for each grid cell before solving the matrix equation.
A similar treatment of irradiation has been described recently by \citet{2013A&A...549A.124B}, for a multi-frequency
implementation see \citet{2010A&A...511A..81K}.
  
\subsection{The matrix solver}
\label{subsec:matrix-solver}
We implemented two different solvers for the matrix equation. The first one uses the 
method of successive over-relaxation (SOR), and as a faster and more flexible solver we use the 
PETSc\footnote{For more information visit the website \url{http://www.mcs.anl.gov/petsc} or have a 
look at \cite{petsc-user-ref}.} library.
From the PETSc library we use the Krylov subspace iterative method and a preconditioner to
solve the matrix equation. For all test cases described we used \mbox{gmres} (Generalized Minimal Residual) 
as iterative methode and \mbox{bjacobi} (Block Jacobi) as preconditioner.
Beside others the convergence of the SOR algorithm
and the PETSc library can be estimated using the following criteria
\begin{equation}
\label{eq:tol}
\left\| \vec{r}^{(k)} \right\| < \max(\epsilon_\mathrm{r} \cdot \left\| \vec{b} \right\|, \epsilon_\mathrm{a})
\end{equation}
where $\vec{b}$ is the right hand 
side of the matrix equation $A\vec{x} =\vec{b}$, $\vec{r}^{(k)}=\vec{b} - A\vec{x}^{(k)}$ is the 
residual vector for the k-th iteration of the solver and $\vec{x}$ is the solution vector (here the radiation energy density).
As norm we used here the $L_2$ norm. The quantities $\epsilon_\mathrm{r}$ and $\epsilon_\mathrm{a}$ are the relative and absolute tolerance,
respectively, and are problem dependent, 
with a common value of $10^{-50}$ for $\epsilon_\mathrm{a}$. For the test cases in section \ref{sec:test_cases} we use relative tolerances
$\epsilon_\mathrm{r}$ between $10^{-5}$ and $10^{-8}$.
The criterion (\ref{eq:tol}) is the 
default one used by the PETSc library. For more information about the convergence test in 
PETSc the reader should refer to section 4.3.2. of \cite{petsc-user-ref}.
The solver performance in a parallel environment is described in section \ref{subsubsec:parallel_scaling}.

\section{Test cases}\label{sec:test_cases}

In order to verify the implemented method, we simulated several test problems and compared the results 
with either corresponding analytical solutions or calculations done with different numerical codes. Most 
of the tests correspond to one-dimensional problems. In order to model those, we have used quasi one-dimensional
domains, with a very long cuboid that has the height $h$, width $w$ and a length $l$. 
The length $l$ is much larger than the width or height, and for simplicity we use $w=h$.
We performed some of the tests in all three implemented coordinate systems (Cartesian, cylindrical and spherical) and in three
different alignments of the cuboid along each coordinate direction. This is done to check whether the geometry factors are correct. In 
the case of a non-Cartesian coordinate system we placed the cuboid at large distances $r$ from the origin
such that the domain approximately describes a Cartesian setup.

We use for all test cases the solver based on the PETSc library with the default iterative solver {\tt gmres} and the pre-conditioner {\tt bjacobi}.

\subsection{Linear diffusion test}\label{sec:linear_diffusion_test}
\begin{figure}
\center
\includegraphics[width=0.48\iswidth]{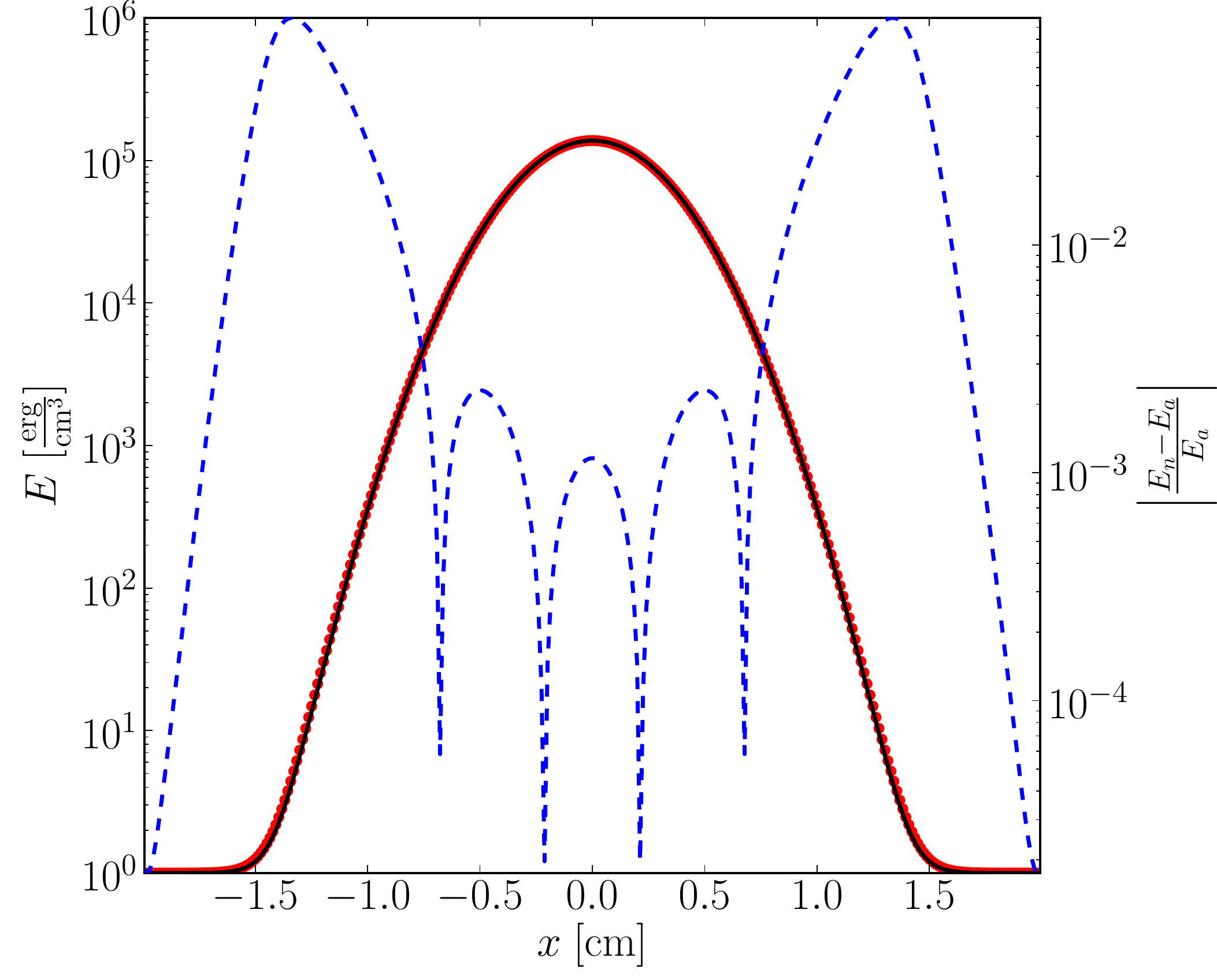}
\caption{Linear diffusion test at the time $t=\unit[4.2 \cdot 10^{-12}]{s}$. 
The simulated (read dots) and the analytical
(black line) solution is plotted. We also plot the solution with the absolute value 
of the relative error (blue dashed line) which belongs to the axis on the right.}
\label{pic:linear_diiffusion_test:comparison}
\end{figure}
\begin{figure}
\center
\includegraphics[width=0.48\iswidth]{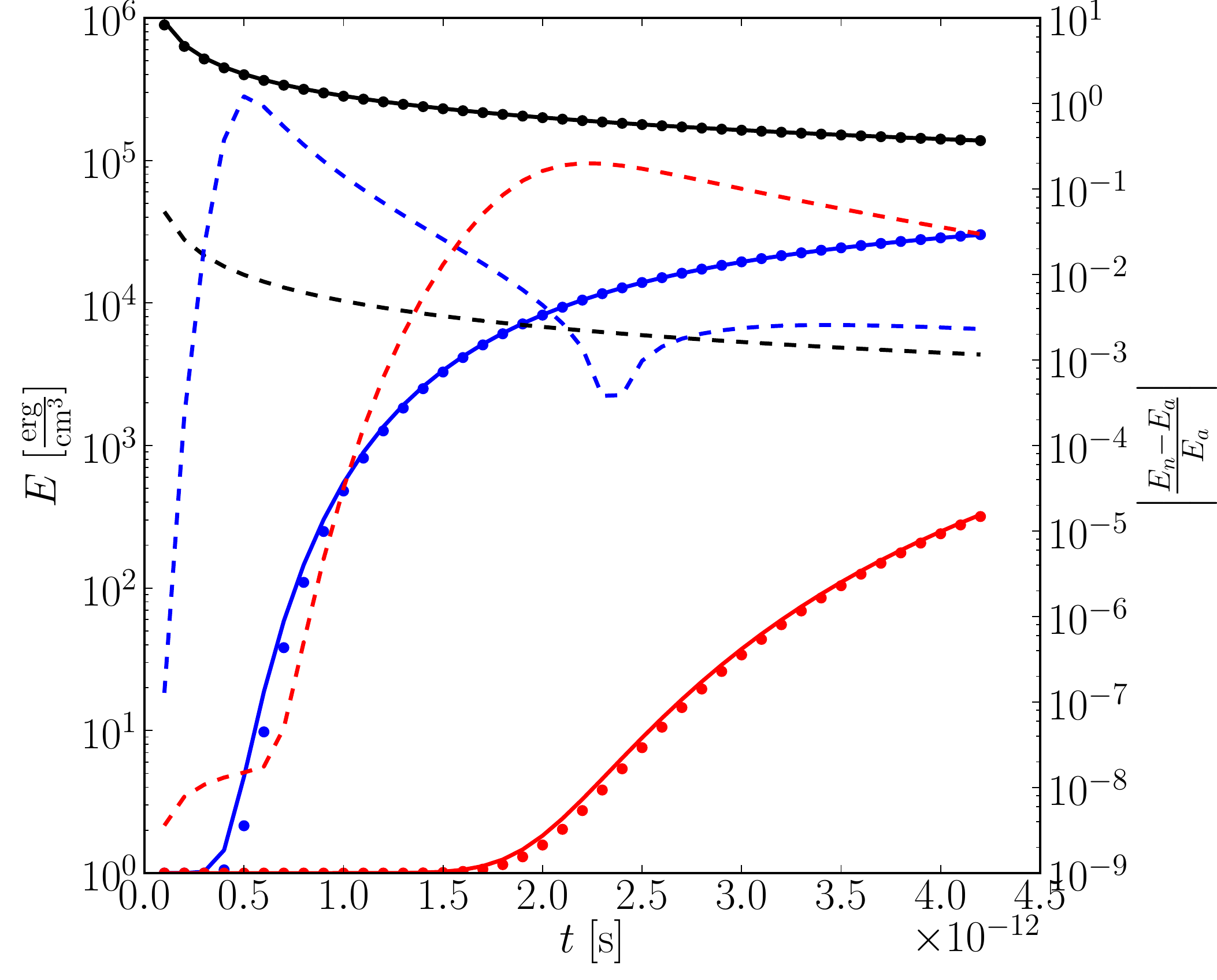}
\caption{
Time evolution for the linear diffusion test from time $t=\unit[0]{s}$
to $\unit[4.2 \cdot 10^{-12}]{s}$ at three different positions at $x = \unit[0]{cm}$ (black lines),
at $x = \unit[0.5]{cm}$ (blue lines) and at $x = \unit[1.0]{cm}$ (red lines).
The dotted lines for each position belong to the simulated solution, the solid lines to the
analytical solution and the dashed lines show the relative error which belong to the axis on the right.
}
\label{pic:linear_diiffusion_test:comparison_time}
\end{figure}
The following test is adapted from \cite{2011A&A...529A..35C}. The initial profile of the radiation 
energy density is set to a delta function which is then 
evolved in time and compared to the analytical one dimensional solution.
We perform this test in all implemented coordinate systems (Cartesian, cylindrical
and spherical coordinates) as described above, which results in nine different simulations.
The used domain is quasi one-dimensional and the equations of hydrodynamics are not solved in this test.
Only the radiation diffusion equation
\begin{equation}
  \pdl{E}{t} = \nabla \cdot \left( \frac{c \lambda}{\kappa_\mathrm{R} \rho} \nabla E \right) 
\label{eqn:linear_diffusion_test:radiation_diffusion_equation}
\end{equation}
is solved which we obtain from equations \eqref{eqn:energy_coupling}
by setting $\kappa_\mathrm{P}=0$. An analytical solution to 
equation \eqref{eqn:linear_diffusion_test:radiation_diffusion_equation} can be calculated in the 
one dimensional case with a constant flux-limiter $\lambda=\frac{1}{3}$ and a constant product of the
Rosseland opacity and density, here we set $\kappa_\mathrm{R} \rho = \unit[1]{cm^{-1}}$. The equation to solve is then given by
\begin{equation}
\pdl{E(x,t)}{t} = \frac{c}{3} \pdld{E(x,t)}{x} 
\label{equation:linear_diffusion_test:radiation_diffusion_1D}
\end{equation}
with solution
\begin{equation}
E(x,t) = \frac{\tilde{E}_0}{\sqrt{\frac{4}{3} c \pi t}}e^{-\frac{3x^2}{4ct}} \,, 
\label{eqn:linear_diffusion_test:solution}
\end{equation}
where $\tilde{E}_0$ is the 
integral over the initial profile of the energy density, $E(x,t=0)$.
Note that in the quasi one-dimensional case
(using a stretched 3D domain) 
$\tilde{E}_0$ has the units $\unit[]{erg \, cm^{-2}}$.

\subsubsection{Setup}

The domain is a cuboid with a length of $\unit[4]{cm}$ and a width and height of 
$\unit[0.04]{cm}$. We used here $301\times3\times3$ grid cells. The initial profile of the 
radiation energy density in the quasi one-dimensional case is set by
\begin{equation}
E_i =\begin{cases}
		\unit[1]{\frac{erg}{cm^3}}, & i = 1,2,\dots,N \text{ with } i\neq \frac{N}{2}\\[0.5em]
		\hspace{10pt}\frac{\tilde{E}_0}{\Delta x},&  i = \frac{N}{2}\\
	\end{cases}
\end{equation}
where $\Delta x$ is the length of a grid cell. For numerical reasons, we have set $E_i$ for $i \neq \frac{N}{2}$ to the 
value $\unit[1]{erg\, cm^{-3}}$ instead of $\unit[0]{erg\, cm^{-3}}$.
This choice is not problematic, since $\tilde{E}_0/\Delta x \gg \unit[1]{erg\, cm^{-3}}$ for our chosen value of 
$\tilde{E}_0 = \unit[10^5]{erg\, cm^{-2}}$. The initial values for pressure and density are 
$p=\unit[1]{g\,cm^{-1}\,s^{-2}}$ and $\rho=\unit[1]{g\,cm^{-3}}$.
Furthermore we use $\kappa_\mathrm{R}=\unit[1]{cm^2\,g^{-1}}$ for the Rosseland opacity.
All boundary conditions are set to periodic 
except for the boundary conditions at the beginning and end of the quasi one-dimensional domain, 
which are set to outflow. For the matrix solver we used a relative tolerance of $\epsilon_r=10^{-8}$.
The simulation starts at $t=\unit[0]{s}$ with an constant time 
step of $\Delta t = \unit[1 \cdot 10^{-14}]{s}$ and stops at $t=\unit[4.2 \cdot 10^{-12}]{s}$.

\subsubsection{Results}

The numerical solution $E_\mathrm{n}$ and the analytical solution $E_\mathrm{a}$ from equation 
\eqref{eqn:linear_diffusion_test:solution} are plotted 
in the figures \ref{pic:linear_diiffusion_test:comparison} and
\ref{pic:linear_diiffusion_test:comparison_time} together with the absolute value of the relative error.
In figure \ref{pic:linear_diiffusion_test:comparison} the radiation energy density is plotted against the
position at the time $t=\unit[4.2 \cdot 10^{-12}]{s}$. The 
relative error in the relevant range from $\unit[-1]{cm}$ to $\unit[1]{cm}$ is always below one percent. 
In figure \ref{pic:linear_diiffusion_test:comparison_time} the time 
evolution from $t=\unit[0]{s}$ to $\unit[4.2 \cdot 10^{-12}]{s}$ is shown for the positions $x=\unit[\{0,0.5,1.0\}]{cm}$
 coded in the colors black, blue and red, respectively.
The results shown in this figure depend strongly on the position. For the 
position $x=\unit[0]cm$ the error is, for all times later than $t=\unit[4\cdot10^{-13}]{s}$, below one percent 
and decreases with time. For the other positions,
the behaviour is different. The relative error rises and after a while it decreases. This behaviour can be explained by looking at
figure \ref{pic:linear_diiffusion_test:comparison}. The error is higher at the diffusion front.
This region moves with time and causes the effect for the other 
positions. The test shows that the time evolution of the radiation energy density is 
reproduced correctly. As described, this test was performed in different coordinate systems and orientations, with the same results.

\subsection{Coupling test}\label{sec:coupling_test}

\begin{figure}
\center
\includegraphics[width=0.415\iswidth]{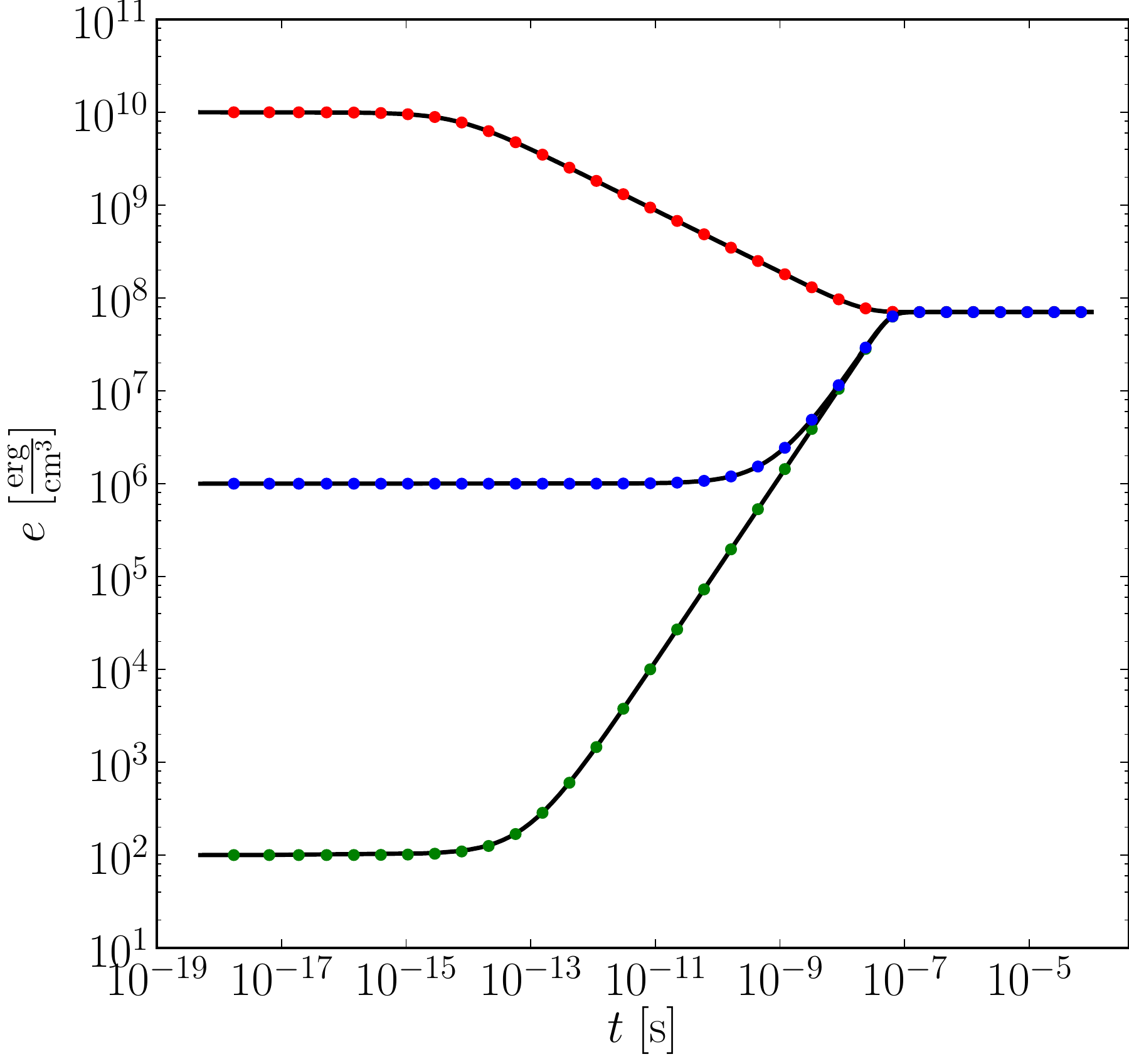}
\caption{Coupling test from $t=\unit[10^{-20}]{s}$ to $t=\unit[10^{-4}]{s}$ with three different 
initial gas energy densities. The reference solution (black lines) and the simulated results for 
the initial energy density $e_0=\unit[10^{10}]{erg \, cm^{-3}}$ (red dots), 
$e_0=\unit[10^{6}]{erg \, cm^{-3}}$ (blue dots) and $e_0=\unit[10^{2}]{erg \, cm^{-3}}$ 
(green dots) are plotted.}
\label{pic:couling_test:comparison}
\end{figure}

The purpose of this test from \cite{2001ApJS..135...95T} is to check the coupling between radiation 
and the fluid. For this purpose we simulate a stationary fluid which is initially out of thermal 
equilibrium. In this simulation the radiation energy density is the dominant energy which is 
constant over the whole simulation. The system of equations \eqref{eqn:energy_coupling} decouples in 
this case and, in addition, it is not necessary to solve the matrix equation for $E$. 
By setting $\sigma_\mathrm{P} = \kappa_\mathrm{P} \rho$ 
and $T = \frac{p}{\rho} \frac{\mu \moH}{k_\mathrm{B}}$ from eq. \eqref{eqn:convert_temperature_to_pressure} with the assumption that 
$\sigma_\mathrm{P}$ and $\rho$ are constant, we can rewrite the thermal energy equation of the system \eqref{eqn:energy_coupling} as  
\begin{equation}
\frac{\id e}{\id t} =\underbrace{c \sigma_\mathrm{P} E}_{C_\mathrm{1}} - \underbrace{c \sigma_\mathrm{P} a_\mathrm{R} 
\left(\frac{\gamma - 1}{\rho} \frac{\mu \moH}{k_\mathrm{B}}\right)^4}_{C_\mathrm{2}} e^4 \,.
\label{eqn:coupling_test}
\end{equation}
With the used approximations, the coefficients $C_\mathrm{1}$ and $C_\mathrm{2}$ are constant. 
The solution to Eq.~(\ref{eqn:coupling_test}) can be calculated analytically in terms of an algebraic equation which would
have to be solved iteratively. Hence, we integrate Eq.~(\ref{eqn:coupling_test}) numerically using a Runge Kutta solver of $4$-th order scheme
with adaptive step size. In the following we refer to this solution as the reference solution.

Information on the expected behaviour of the solution can be obtained directly from the differential equation.
It is clear that in the final equilibrium state (with $\frac{\id e}{\id t}=0$) the gas temperature has to be equal to the
radiation temperature $T=\sqrt[4]{\frac{E}{a_\mathrm{R}}}$, 
thus the final gas energy density will be 
\begin{equation}
e_\mathrm{final} = \left( \frac{C_\mathrm{1}}{C_\mathrm{2}} \right)^{\frac{1}{4}}\,.
\end{equation}
If the initial gas energy density $e_0$ is much lower than $e_\mathrm{final}$, we 
can neglect the second term in eq.~\eqref{eqn:coupling_test} at the beginning, thus $e (t) = C_\mathrm{1}\,t + e_0$. 
The corresponding coupling time can be estimated to 
\begin{equation}
\tau = \frac{e_\mathrm{final} - e_0}{C_\mathrm{1}}\,. \label{eqn:coupling_test:coupling_time_low_e0}
\end{equation}
On the other hand, if $e_0 \gg e_\mathrm{final}$, we can neglect the first term in eq. \eqref{eqn:coupling_test} and derive 
\begin{equation}
e (t) \propto \left( C_\mathrm{2}\,t \right)^{-\frac{1}{3}} \quad  \mathrm{and} \quad \tau = \frac{1}{e_\mathrm{final}^3\,C_\mathrm{2}}\,.\label{eqn:coupling_test:coupling_time_high_e0}
\end{equation}

\subsubsection{Setup}

The computational domain is identical to that of the linear diffusion test in section 
\ref{sec:linear_diffusion_test}. For the grid we use a resolution of $25\times3\times3$ grid 
cells. As before we do not solve the equations of hydrodynamics and the boundary conditions are 
quite simple. All boundaries are set to periodic boundary conditions. The constants we used are set 
to: radiation energy density $E=\unit[10^{12}]{erg \, cm^{-3}}$, density $\rho = 
\unit[10^{-7}]{g \, cm^{-3}}$, opacity $\sigma_\mathrm{P} = \unit[4\cdot 10^{-8}]{cm^{-1}}$, mean 
molecular weight $\mu=0.6$ and the ratio of specific heats $\gamma = 5/3$.
The simulations starts at $t=\unit[0]{s}$ with an initial time step of $\Delta t = \unit[10^{-20}]{s}$ and evolves 
until $t=\unit[10^{-4}]{s}$. After each step the time step is increased by 1\% in order to 
speed-up the computation.
The simulation is done 
with three different initial gas energy densities, $e_0=\unit[10^{10}]{erg \, cm^{-3}}$, 
$e_0=\unit[10^{6}]{erg \, cm^{-3}}$ and $e_0=\unit[10^{2}]{erg \, cm^{-3}}$.

\subsubsection{Results}
Figure \ref{pic:couling_test:comparison} shows the numerical gas energy density and the reference solution plotted against
time for the three different initial values of $e$. The agreement of both results is excellent for all initial values.
From the figure we see that in the limit of small and large initial $e_0$, we find exactly the behaviour
as predicted by the estimates for eq.~\eqref{eqn:coupling_test}.
The analytic estimates for the coupling time $\tau$ from equation \eqref{eqn:coupling_test:coupling_time_low_e0} agrees very well with our results. The estimate for $e_0=\unit[10^{2}]{erg \, cm^{-3}}$ is $\tau=\unit[5.88\cdot 10^{-8}]{s}$ and for $e_0=\unit[10^{6}]{erg \, cm^{-3}}$ we calculated $\tau=\unit[5.78\cdot 10^{-8}]{s}$. 
In the case that $e_\mathrm{final} < e_0$ the estimate in eq.~\eqref{eqn:coupling_test:coupling_time_high_e0} is approximate $\tau=\unit[5.88\cdot 10^{-8}]{s}$.
We have to mention here that this test verifies primarily the correctness of equation \eqref{eqn:calculation_temperature}.
As in the linear diffusion test, this test was performed in three different coordinate systems in different orientations, with the same results.

\subsection{Coupling test with irradiation}\label{sec:coupling_test_irradiation}

\begin{figure}
\center
\includegraphics[width=0.415\iswidth]{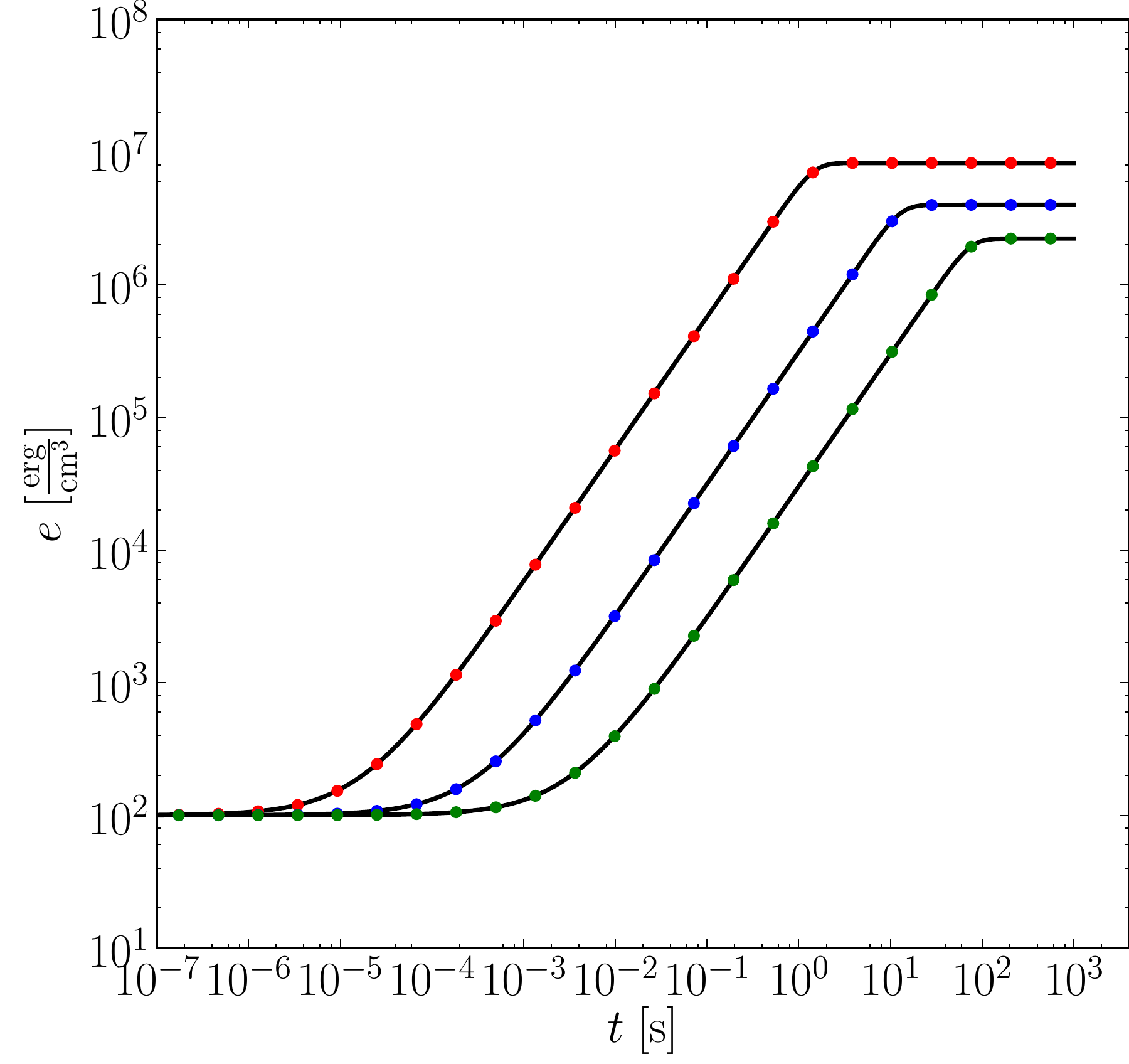}
\caption{Coupling test with enabled irradiation from $t=\unit[10^{-7}]{s}$ to $t=\unit[10^{3}]{s}$ 
at three different distances $d$ from the inner boundary of the domain. The reference solution (black lines) and the simulated 
results for the energy density $e=\unit[10^{2}]{erg \, cm^{-3}}$ are plotted at 
distance $d=\unit[0]{cm}$ (red dots), $d=\unit[3\cdot10^4]{cm}$ (blue dots) and 
$d=\unit[3\cdot10^5]{cm}$ (green dots).}
\label{pic:couling_test:comparison_irradiation}
\end{figure}

\begin{figure}
\center
\includegraphics[width=0.415\iswidth]{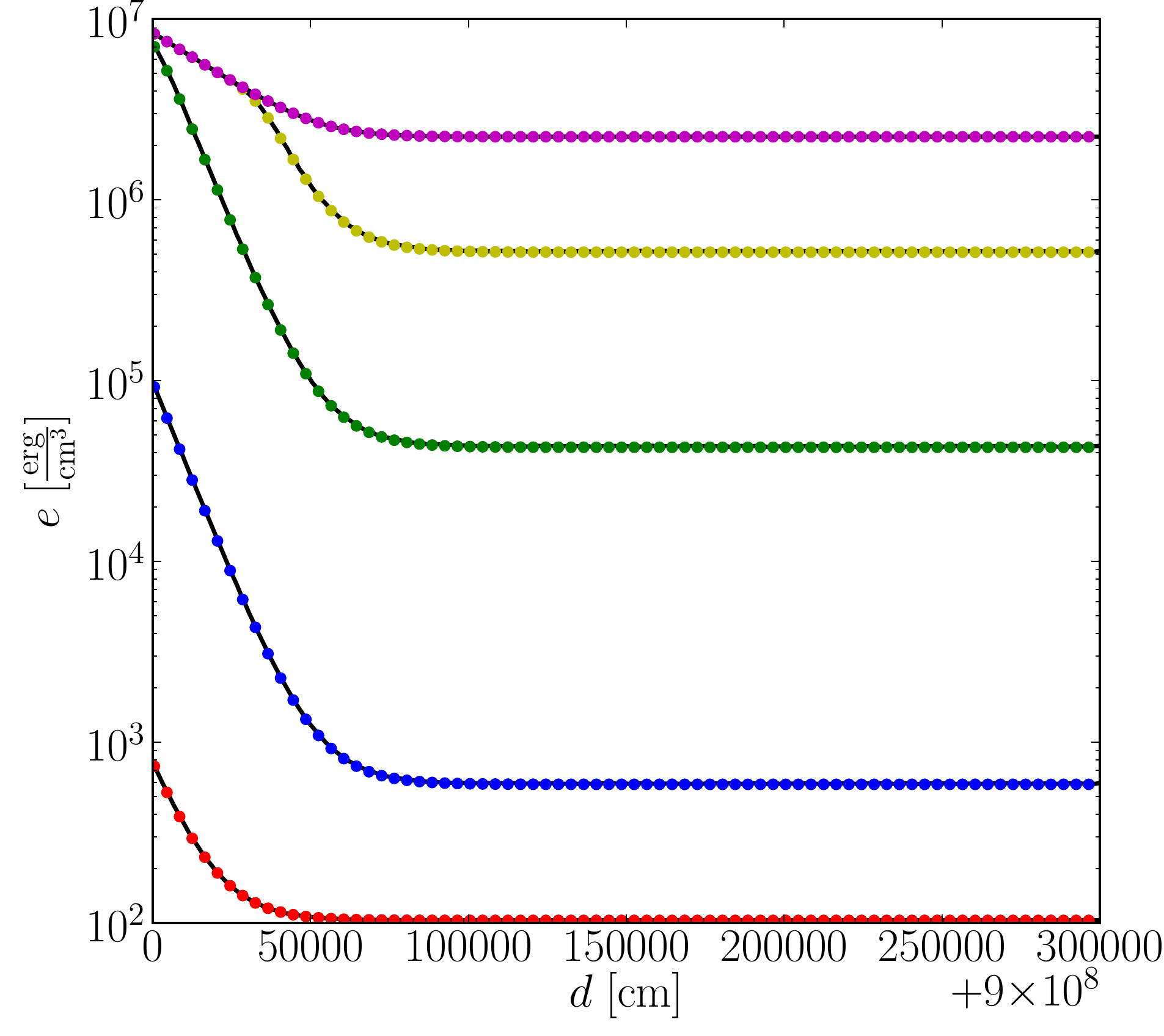}
\caption{Radial dependency of the gas energy density for the coupling test with enabled irradiation. The 
gas energy density is plotted at different times, $t=\unit[1.12\cdot 10^{-4}]{s}$ (red dots),
$t=\unit[1.62\cdot 10^{-2}]{s}$ (blue dots), $t=\unit[1.43]{s}$ (green dots), $t=\unit[1.72\cdot 10^{1}]{s}$ (yellow dots)
and $t=\unit[9.18\cdot 10^{2}]{s}$ (magenta dots). The dots represent the numerical solution and the solid black lines the 
reference solution. The position $d$ is again measured relative to the inner boundary of the quasi one-dimensional domain.}
\label{pic:couling_test:comparison_irradiation_r}
\end{figure}

This test is in its basic setup the same as that from section \ref{sec:coupling_test}, but with 
irradiation enabled, i.e. equation \eqref{eqn:fluid_internal_energy_coupling_irradiation} is solved 
instead of the second equation in \eqref{eqn:energy_coupling}. As described in section \ref{sec:irradiation}, 
irradiation is limited to spherical coordinates, which we use for this test.
With the same assumptions as in section \ref{sec:coupling_test}
, i.e., that $\sigma_\mathrm{P}$ and $\rho$ are constant and with the definitions for
$\sigma_\mathrm{P}$, $e$, $p$ as well as for $T$, it is possible to
rewrite $S$ from equation \eqref{eqn:S} to
\begin{equation}
S(r)= \frac{3 \sigma T_\star^4 R^2_\star e^{-\sigma_\mathrm{P} \left( r - r_\mathrm{0}\right)}\left(1-e^{-\sigma_\mathrm{P} 
\Delta r}\right)}{\left(r+\Delta r\right)^3 - r^3} \,,
\end{equation}
and obtain for equation \eqref{eqn:fluid_internal_energy_coupling_irradiation}
\begin{equation}
\frac{\id e}{\id t}
=\underbrace{S(r) + c \sigma_\mathrm{P} E}_{C_\mathrm{1}(r)} - \underbrace{c \sigma_\mathrm{P} a_\mathrm{R} \left(\frac{\gamma - 
1}{\rho} \frac{\mu \moH}{k_\mathrm{B}}\right)^4}_{C_\mathrm{2}} e^4 \,.
\end{equation}
The reference solution is computed in the same way as before although it now depends on
the distance $r$ from the star. The quasi one-dimensional domain starts at 
$r=\unit[9000 \cdot 10^5]{cm}$ and ends at $r=\unit[9003 \cdot 10^5]{cm}$ and we use $300\times 3 
\times 3$ grid cells. The domain size in $\theta$ and $\phi$ direction was chosen in a way such 
that the grid cells are nearly quadratic. For the simulation we use a constant radiation energy 
density of $E=\unit[10^{-2}]{erg \, cm^{-3}}$, a density of $\rho=\unit[10^{-5}]{g \, cm^{-3}}$
, a Rosseland opacity of $\kappa_{\rm R}=\unit[10]{cm^2\, g^{-1}}$ and a Planck opacity of $\kappa_\mathrm{P} = \kappa_\mathrm{R}$
which corresponds to $\sigma_\mathrm{P} = \unit[10^{-4}]{cm^{-1}}$. The opacity for the irradiation $\kappa_\star$ is set
to $\kappa_\mathrm{P}$. For the star, the temperature was 
set to $T_\star=\unit[6000]{K}$ and the radius to $R_\star=\unit[8.1 \cdot 10^8]{cm}$.
Additionally we make the assumption that there is no absorption in the region between the surface of
the star and the inner boundary of the computation domain.
Figure \ref{pic:couling_test:comparison_irradiation} shows the gas energy density plotted against time
with an initial gas energy of $e=\unit[10^{2}]{erg \, cm^{-3}}$ at three different 
positions $d=\unit[0]{cm}$, $d=\unit[3\cdot10^4]{cm}$ and $d=\unit[3\cdot10^5]{cm}$ where $d$ is 
measured relative to the inner boundary of the quasi one-dimensional domain.
In figure \ref{pic:couling_test:comparison_irradiation_r} the radial dependency of the gas energy density
is plotted for the same simulation at five different times.
As expected, the results show that the gas 
energy density at a time later than $t=\unit[10^2]{s}$ becomes constant and depends on the 
distance from the star.  The simulated and reference solution show a excellent agreement. 

\subsection{A steady state test}\label{sec:steady_state_test}

\begin{figure}
\center
\includegraphics[width=0.462\iswidth]{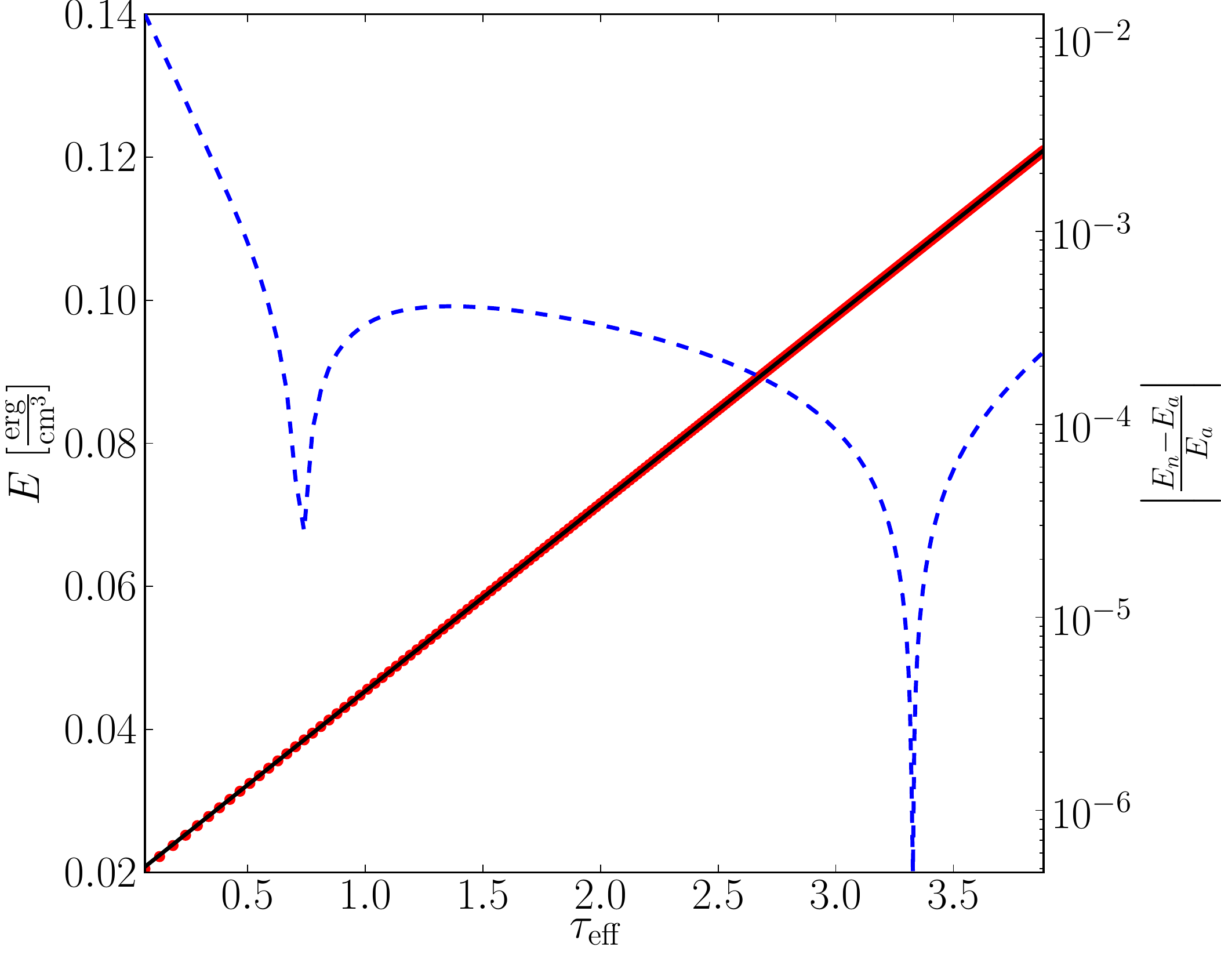}
\caption{Comparison between the numerical (red dots) and the analytical (black line) solution of 
the steady state test after $t=\unit[1200]{s}$. In addition the absolute value of the relative 
error (blue dashed line) is plotted related to the axis on the right side. Note this axis is 
logarithmic.}
\label{pic:steady_state_test:comparison}
\end{figure}

The original version of this test was published in \cite{Flaig11}. We consider a one-dimensional stationary
setup with a given density stratification. 
In the steady state, the time derivatives in the equations \eqref{eqn:energy_coupling} 
vanish and the system is reduced to the following equation for the radiation energy density
\begin{equation}
  0 = \nabla \cdot \, \left( \frac{c \lambda}{\kappa_\mathrm{R} \rho} \nabla E \right) \,.
  \label{eqn:steady_state_test:steady_state_base_equation}
\end{equation}
A further reduction is obtained when we rewrite this equation in one dimension along the z-axis in 
Cartesian coordinates. The equation is then much simpler and can be written as
\begin{equation}
	\frac{\id}{\id z} \left( \frac{c \lambda}{\kappa_\mathrm{R} \rho} \frac{\id}{\id z} E \right) = 0 \,. 
\label{eqn:steady_state_test:1D_steady_state_base_equation}
\end{equation}
In general the expression $\frac{c \lambda}{\kappa_\mathrm{R} \rho}$ is not known analytically for realistic opacities.
In order to circumvent this problem, we define the effective optical depth $\tau_\mathrm{eff}=\int\id 
\tau_\mathrm{eff}=\int_{z_b}^{z_a}\kappa_\mathrm{eff} \rho \id z$ where $z_a$ and $z_b$ are the lower and upper boundaries of the 
quasi one-dimensional domain, respectively, and $\kappa_\mathrm{eff}$ is the effective opacity given by 
$\kappa_\mathrm{eff}=\frac{1}{3}\frac{\kappa_\mathrm{R}}{\lambda}$. By using $\id 
\tau_\mathrm{eff}=\kappa_\mathrm{eff} \rho \id z$, equation 
\eqref{eqn:steady_state_test:1D_steady_state_base_equation} can be rewritten as:
\begin{equation}
	\frac{\id}{\id z} \frac{\id }{\id \tau_\mathrm{eff}} E = 0 \,.
\end{equation}
The solution of this equation is then given by
\begin{equation}
  E = \left(E(\tau_{\mathrm{eff}} = 1) - E(\tau_{\mathrm{eff}} = 0)\right) \tau_{\mathrm{eff}} + 
  E(\tau_{\mathrm{eff}} = 0) \label{eqn:steady_state_test:steady_state_solution}
\end{equation}
where $E(\tau_{\mathrm{eff}} = 0)$ and $E(\tau_{\mathrm{eff}} = 1)$ are the radiation energy density at 
the position where the effective optical depth has the values zero or one, respectively.
Thus, in the static case the radiation energy has a linear dependence on the optical depth $\tau_{\mathrm{eff}}$
for all opacity laws.

\subsubsection{Initial setup}

The domain was chosen to have an arbitrary length of $\unit[300]{cm}$ 
and a width and height of $\unit[3]{cm}$ and $300 \times 3 \times 3$ grid cells were used. 
This test is performed without solving the hydrodynamical equations, instead we solved 
equations \eqref{eqn:energy_coupling} for a fixed density and opacity law, and evolved
the solution, until a stationary state has been reached.
For the radiation boundary conditions, we used boundary 
conditions with fixed values of $E$ at the lower and upper boundary of the domain. At the lower boundary we have chosen 
$E=a_\mathrm{R} T^4$ with a temperature of $T=\unit[2000]{K}$.

Because the stratification is optically thin at the upper boundary we want to allow the radiation to escape freely from the
domain. For this reason we simply set the temperature to a very small value at the upper boundary, here  $T=\unit[10]{K}$. 

All other 
boundary conditions have been set to periodic. The density stratification is given 
by
\begin{equation}
\rho(z) = \rho_0  e^{\frac{1}{2} \left( \frac{z - z_{a}}{ 0.46 \cdot \left( z_{b} - z_{a}\right)}\right)^2} \,.
\end{equation}
The initial 
temperature profile can be chosen randomly in principle, but in order to speed up the computation 
we used a linear temperature profile starting at $z_a$ with $T=\unit[2000]{K}$ and ending at $z_b$ 
with $T=\unit[10]{K}$. From this temperature profile we assigned pressure values using equation 
\eqref{eqn:convert_temperature_to_pressure}. The radiation energy density $E$ inside the domain is also 
set using the gas temperature profile and $E=a_\mathrm{R} T^4$. The ratio of specific heats and the mean molecular 
weight are set to $\gamma=1.43$ and $\mu = 0.6$, respectively. As flux-limiter we have chosen equation 
\eqref{eqn:fluxlimiter_minerbo}, for the Rosseland mean opacity $\kappa_{\rm R}$ we use data from 
\cite{1985prpl.conf..981L}, and the Planck mean opacity is set to $\kappa_{\rm P} = \kappa_{\rm R}$.
The initial time step is $\Delta t = \unit[0.3]{s}$ and it is increased slightly with time in order to speed-up
the computation and to keep the number of iterations done by the matrix solver nearly constant. 
This simulation was preformed with a relative tolerance of $\epsilon_r=10^{-6}$ for the matrix solver.

\subsubsection{Results}

A steady state is reached approximately after $t=\unit[1200]{s}$. In figure \ref{pic:steady_state_test:comparison},
we plot the radiation energy density against the effective optical depth $\tau_\text{eff}$ from our 
numerical solution (red dots) together with the analytical solution from equation 
\eqref{eqn:steady_state_test:steady_state_solution}.  
The parameters $E(\tau_{\mathrm{eff}} = 1) - E(\tau_{\mathrm{eff}} = 0)$ and $E(\tau_{\mathrm{eff}} = 0)$
have been obtained by fitting equation 
\eqref{eqn:steady_state_test:steady_state_solution} to the numerical solution. 
We have to note here that $E(\tau_{\mathrm{eff}} = 0)$ is determined by interpolation between ghost cells and active
cells near the upper boundary $z_\mathrm{b}$. Hence, the radiation temperature in the active region can be much larger
than $\unit[10]{K}$, a value was specifically chosen to be very small.
We also plot the absolute value of the relative error $\left|(E_n -E_a) / E_a\right|$. The 
results from the simulation agree very well with the analytical prediction. As we can see from 
figure \ref{pic:steady_state_test:comparison}, the largest deviation from the analytical solution is 
at small values of $\tau_\mathrm{eff}$ with an relative error around one percent. As the linear diffusion and the coupling test, 
this test was performed in all three coordinate systems and in different orientations, with the same results.

\subsection{Radiation shock}\label{sec:radiation_shock}

\begin{figure}
\center
\subfigure[Subcritical shock]{\label{pic:radiation_shock_test:subcritical}
  \includegraphics[width=0.415\iswidth]{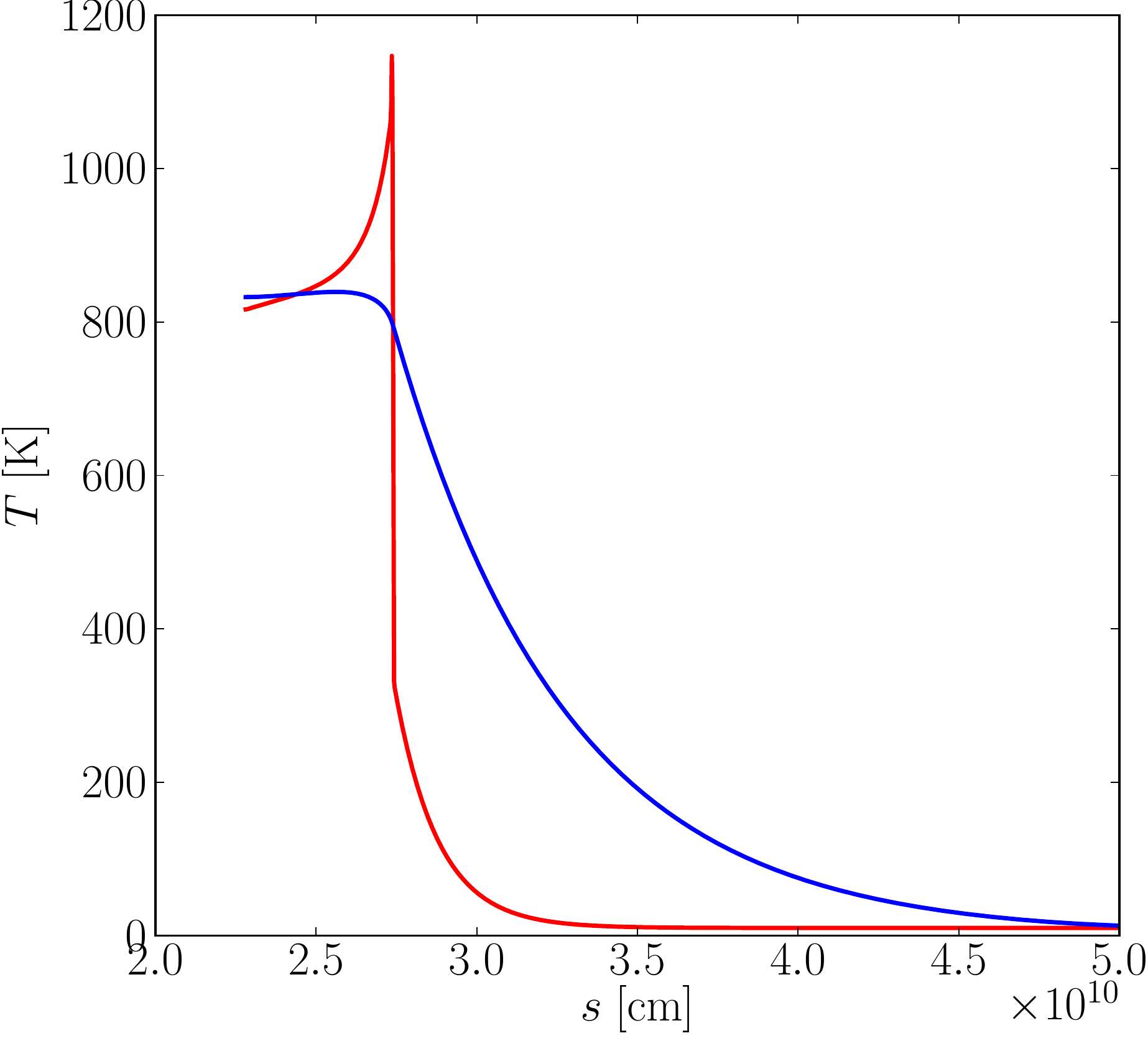}}
\subfigure[Supercritical shock]{\label{pic:radiation_shock_test:supercritical}
  \includegraphics[width=0.415\iswidth]{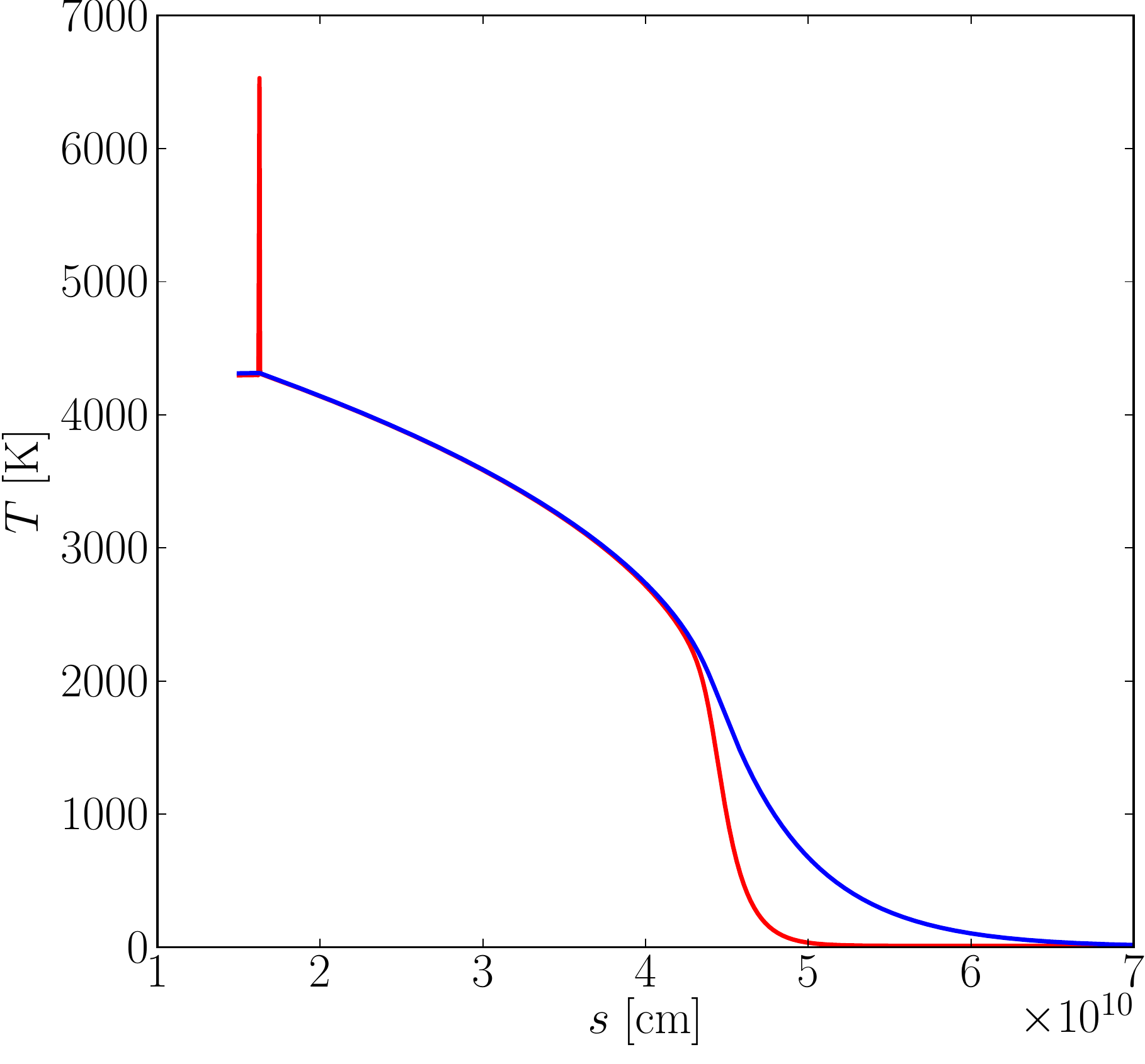}}
\caption{Sub- and supercritical shock test. In both cases we plot the radiation temperature (blue 
line) and the gas temperature (red line) against $s = z - v\cdot t$ where $z$ is the position along 
the quasi one-dimensional domain and $v$ the piston velocity. The subcritical shock 
\subref{pic:radiation_shock_test:subcritical} is shown at time $t = \unit[3.8\cdot10^4]{s}$ and 
the supercritical shock \subref{pic:radiation_shock_test:supercritical} at $t = \unit[7.5\cdot10^3]{s}$.}
\label{pic:radiation_shock_test}
\end{figure}

In this section we extend the previous tests and solve now the full equations of 
hydrodynamics and radiation transport simultaneously, testing the complete new module within \PLUTO
environment. 

\subsubsection{Initial setup}

Following a set-up from \cite{1994ApJ...424..275E}, a shock is generated in a
quasi one-dimensional domain.
This test case is more complex than the ones
before, and it is not possible to derive an analytical solution. Instead we compare our 
results with the simulations of \cite{2011A&A...529A..35C}. 
The computational domain is chosen to have a length of $\unit[7 \cdot 10^{10}]{cm}$ and a width and height of 
$\unit[3.418 \cdot 10^{7}]{cm}$ with $2048 \times 4 \times 4$ grid cells. The initial 
density and temperature are set to $\rho = \unit[7.78 \cdot 
10^{-10}]{g \, cm^{-3}}$ and $T = \unit[10]{K}$. The initial radiation energy 
density is set by the equation $E = a_\mathrm{R} T^4$. For the flux-limiter we employ
the Minerbo-formulation according to eq.~\eqref{eqn:fluxlimiter_minerbo}, and for the opacity we use
$\kappa_\mathrm{R} \cdot \rho = \kappa_\mathrm{P} \cdot \rho = \unit[3.1 \cdot 10^{-10}]{cm^{-1}}$.
Furthermore the ratio of specific heats is set to $\gamma=7/5$
and the mean molecular weight to $\mu = 1$, in analogy to \cite{2011A&A...529A..35C}. The time step is computed through 
the CFL condition of \PLUTO\, for wich we assume a value of $0.4$. For the solver we took the not so accurate but robust
tvdlf which uses a simple Lax-Friedrichs scheme. For generating the radiative shock, the following boundary conditions are used: 
in the direction of the shock propagation, we employ a reflective boundary condition at the lower boundary and a 
zero-gradient at the upper boundary of the domain. The remaining boundaries are set to periodic.
For the relative tolerance used by the matrix solver we have chosen a value of $\epsilon_r=10^{-5}$.
The shock is generated by 
applying an initial velocity $v$ to the gas. The velocity is directed towards the reflecting 
boundary condition which acts as a wall. The shock propagates then from the wall back into the 
domain. Depending on the velocity, the shock is sub- or supercritical, i.e., the temperature behind the shock front is 
larger or equal than the temperature upstream (in front of the shock front), respectively.
In this test we simulate both cases: the subcritical shock with a 
velocity of $v = \unit[6\cdot 10^5]{cm \, s^{-1}}$ and the supercritical shock with $v = 
\unit[20\cdot 10^5]{cm \, s^{-1}}$. 

\subsubsection{Results}

For a better comparison with the results of simulations, 
where the material is at rest and a moving piston causes the shock, we introduce the quantity $s$. 
This quantity is given by the relation $s = z - v \cdot t$ where $z$ is the position along the 
quasi one-dimensional domain. Note that this quantity is called 
$z$ in \cite{2011A&A...529A..35C}. Fig. \ref{pic:radiation_shock_test} shows the radiation temperature (blue line) and
the gas temperature (red line) against the previously defined quantity $s$ for both the subcritical (at $t = 
\unit[3.8\cdot10^4]{s}$) and supercritical case (at $t=\unit[7.5\cdot10^3]{s}$). 
In the supercritical case the pre- and post-shock
gas temperature are equal, as expected. In the subcritical case these temperatures can be estimated 
analytically \citep{1994ApJ...424..275E,1984oup..book.....M,2011A&A...529A..35C}. In table \ref{tab:radiation_shock}, the 
analytical estimates and the numerical values from our simulations and the results from \cite{2011A&A...529A..35C} are shown together.
Here $T_2$ is the post-shock temperature, $T_{-}$ the pre-shock temperature and $T_{+}$ the spike temperature. 
In the equations, $R_{\rm G} =\frac{k_\mathrm{B}}{\mu \moH}$ is the perfect gas constant, 
$\sigma_\mathrm{SB}=\frac{c a_\mathrm{R}}{4}$ the Stefan-Boltzmann constant,
and $u$ is the velocity of the shock relative to the upstream material (or vice versa) in our case $u=\unit[7.19\cdot 10^5]{cm \, s^{-1}}$.

\begin{table}[h!]
\center
\begin{tabular}{ l | p{1.3 cm} | p{1.3cm}| p{1.2 cm} }
& analytical estimate &  numerical solution & Commer{\c c}on et al. \\
\hline
$T_2 \approx\frac{2(\gamma - 1) u^2}{R_{\rm G} (\gamma + 1)^2}$& $\sim\phantom{1}\unit[865]{K}$ & $\phantom{1}\unit[816.6]{K}$ & $\phantom{1}\unit[825]{K}$ \\
$T_{-} \approx \frac{\gamma - 1}{\rho u R_{\rm G} }\frac{2 \sigma_\mathrm{SB} T_2^4 }{\sqrt{3}}$ & $\sim\phantom{1}\unit[315]{K}$ &$\phantom{1}\unit[331.9]{K}$& $\phantom{1}\unit[275]{K}$ \\
$T_{+} \approx T_2 + \frac{3-\gamma}{\gamma + 1} T_{-}$& $\sim\unit[1075]{K}$ &$\unit[1147.1]{K}$& $\unit[1038]{K}$ \\
\end{tabular}
\caption{Comparison of the results from the radiation shock test with analytical estimates and the results
from \cite{2011A&A...529A..35C} for the pre-shock $T_{-}$ and post-shock $T_2$ gas temperature as well as the spike temperature $T_{+}$. }
\label{tab:radiation_shock}
\end{table}
The results agree in general with the analytical estimates and the results from \cite{2011A&A...529A..35C}.
The analytical estimate for the post-shock temperature is higher than the numerical results with both codes.
We have to note here that the analytical estimate depends on $u$ and differs therefore from the values given
in \cite{2011A&A...529A..35C}. The pre-shock and spike temperatures agree reasonably well with the analytical estimates in our
simulations but are higher than the results from \cite{2011A&A...529A..35C}.
The differences of our numerical solution to the analytical estimates might due to the fact that we ignored the advective terms in the
radiation energy density in eq.~(\ref{eqn:radiation_energy_equation_coupled}) that may play a role in this dynamic situation.
Additionally, it is noteworthy that the position of the shock front is very well reproduced.
This test was performed in Cartesian coordinates.

\subsection{Accretion disc}\label{sec:accretion_disc}

The goal of this last test is to compare the results of different codes on a more complex two-dimensional physical problem
that involves the onset of convective motions.
For this purpose we model a section of an internally
heated, viscous accretion disc in spherical coordinates $(r,\theta,\phi)$ where $r$ is the distance to the centre 
of the coordinate system, $\theta$ the polar angle measured from the $z$-axis in cylindrical coordinates 
and $\phi$ the azimuth angle. The setup follows the standard disc model used in \citet{2009A&A...506..971K}. 
The tests proceed in two steps. In a first setup we reduce the complexity of the problem and consider 
a static problem, i.e., without solving the equations of hydrodynamics.
This will demonstrate that the equilibrium between viscous heating and radiative cooling is treated correctly
in our implementation. In the second setup we consider the full hydrodynamic problem
and study the onset of convection in discs.

\subsubsection{The initial setup}\label{subsubsec:ad-init}
For both, the static and the dynamical case we use the same initial setup.
The radial extent ranges from $r_\text{min} = 0.4$ to $r_\text{max} = 2.5$, where all lengths are given in units
of the semi-major axis of Jupiter $a_\text{jup}= \unit[5.2]{AU}$. In the vertical direction the domain extends
from $\theta_\text{min} = \unit[83]{^\circ}$ to $\theta_\text{max} = \unit[90]{^\circ}$ and in $\phi$ direction from 
$\phi_\text{min} = \unit[0]{^\circ}$ to $\phi_\text{max} = \unit[360]{^\circ}$. In the 
three coordinate directions $(r,\theta,\phi)$ we use $256\times 32 \times 4$ grid cells.
The disc aspect ratio $h$ is set to $h=\frac{H}{s}= \unit[0.05]{}$ where $s=r \sin \theta$ describes the (radial)
distance from the $z$-axis in cylindrical coordinates, and $H$ is the disc's vertical scale height. The viscosity $\nu$ is set to a
value of $\nu=\unit[10^{15}]{cm^2 \, s^{-1}}$, and the mean molecular weight to $\mu=\unit[2.3]{}$. 
For the ratio of specific heats we have used different values, as specified below.
The density stratification can be obtained
from vertical hydrostatic equilibrium, assuming a temperature that is constant on cylinders, $T = T(s)$. 
It follows \citep{2006ApJ...652..730M}
\begin{equation}
  \rho(r,\theta) = \rho_0 \cdot s^{-1.5} \exp\left(\frac{\sin{\theta} - 1}{h^2}\right)
\end{equation}
where the quantity $\rho_0$ was chosen such that the total mass of the disc
is $M_\text{disc} = 0.01 \cdot M_\star$, where 
$M_\star$ is the mass of the central star of the system which is set to the mass of the sun, $M_\star = M_\odot$.
The mass within the computational domain 
is then $1/2 M_\text{disc}$ because we only compute the upper half of the disc.
The radial variation leads to a surface density profile of $\Sigma \propto r^{-1/2}$, which
is the equilibrium profile for constant viscosity, and vanishing mass flux through the disc.
The pressure $p$ is set by the isothermal relation $p = \rho c_\mathrm{s}^2$, with the speed
of sound $c_\mathrm{s} = H \Omega_\mathrm{K}$ and the Keplerian angular velocity
\begin{equation*}
\Omega_\mathrm{K} = \sqrt{\frac{G M_\star}{s^3}}\,,
\end{equation*}
with the gravitational constant $G$.
The temperature can be computed through equation \eqref{eqn:convert_temperature_to_pressure} and results in $T=\frac{\mu \moH}{k_\mathrm{B}}\frac{p}{\rho}$.
The initial velocities are set to zero except for the angular velocity $v_\phi$ which is set to
\begin{equation*}
v_\phi = \sqrt{\frac{\left(1 - 2 h^2\right) G M_\star}{s}} \,.
\end{equation*}
For the Rosseland mean opacity $\kappa_{\rm R}$ we use data from 
\cite{1985prpl.conf..981L}, and the Planck mean opacity is set to $\kappa_{\rm P} = \kappa_\mathrm{R}$.
The displayed simulations have been performed in the rotating frame in which the coordinate system rotates
with the constant angular velocity of 
$\Omega_\mathrm{K}$ at $a_\text{jup}$, but for non-rotating systems identical results are obtained.
As before the radiation energy density is initialised to $E=a_\mathrm{R} T^4$.

For density, pressure and radial velocity we apply reflective radial boundary conditions and the angular velocity is set to the Keplerian values. In the azimuthal direction periodic boundary conditions are used for all variables. In the vertical direction we apply an equatorial symmetry and reflective boundary condition for $\theta_{min}$.
The radiation 
boundary conditions are set to reflective for the $r$ direction (both lower and upper),
in $\theta$-direction we use a fixed value of $E=a_\mathrm{R} T^4$ with $T=\unit[5]{K}$ at $\theta_{min}$ 
(which denotes the disc surface), and a symmetric boundary condition holds at the disc's midplane $\theta_{max}$. 
For the $\phi$-direction we use periodic boundary conditions.

In both cases we used for the matrix solver a relative tolerance of $\epsilon_r=10^{-8}$.
In the simulation with hydrodynamics we use the Riemann-solver hllc\footnote{Harten, Lax, Van Leer approximate Riemann Solver with the contact discontinuity}.

\begin{figure}
\center
\subfigure[$t=10 \text{ orbits}$]{\label{pic:accretion_disc_test:10orbits}
  \includegraphics[width=0.46\iswidth]{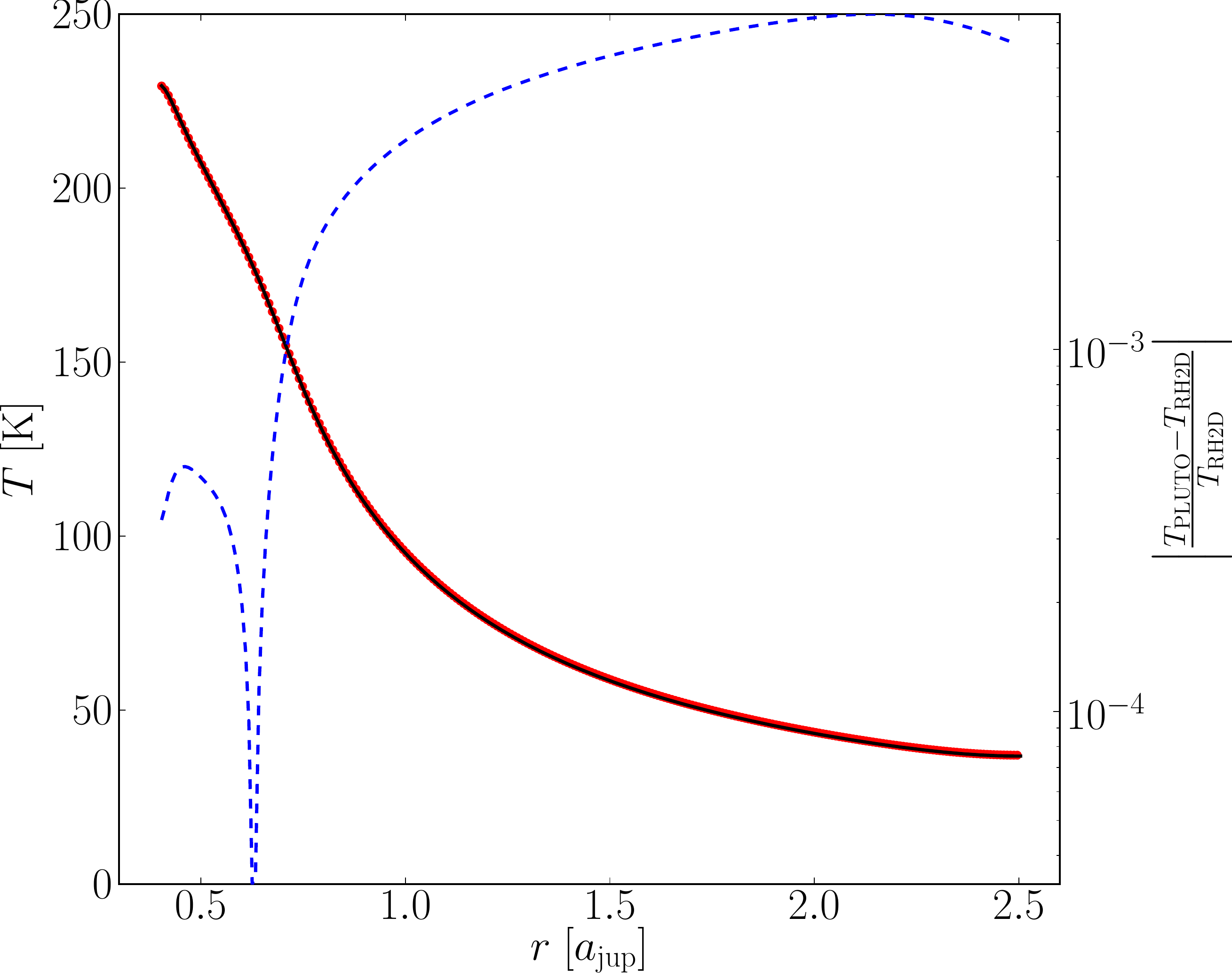}}
\subfigure[$t=100 \text{ orbits}$]{\label{pic:accretion_disc_test:100orbits}
 \includegraphics[width=0.46\iswidth]{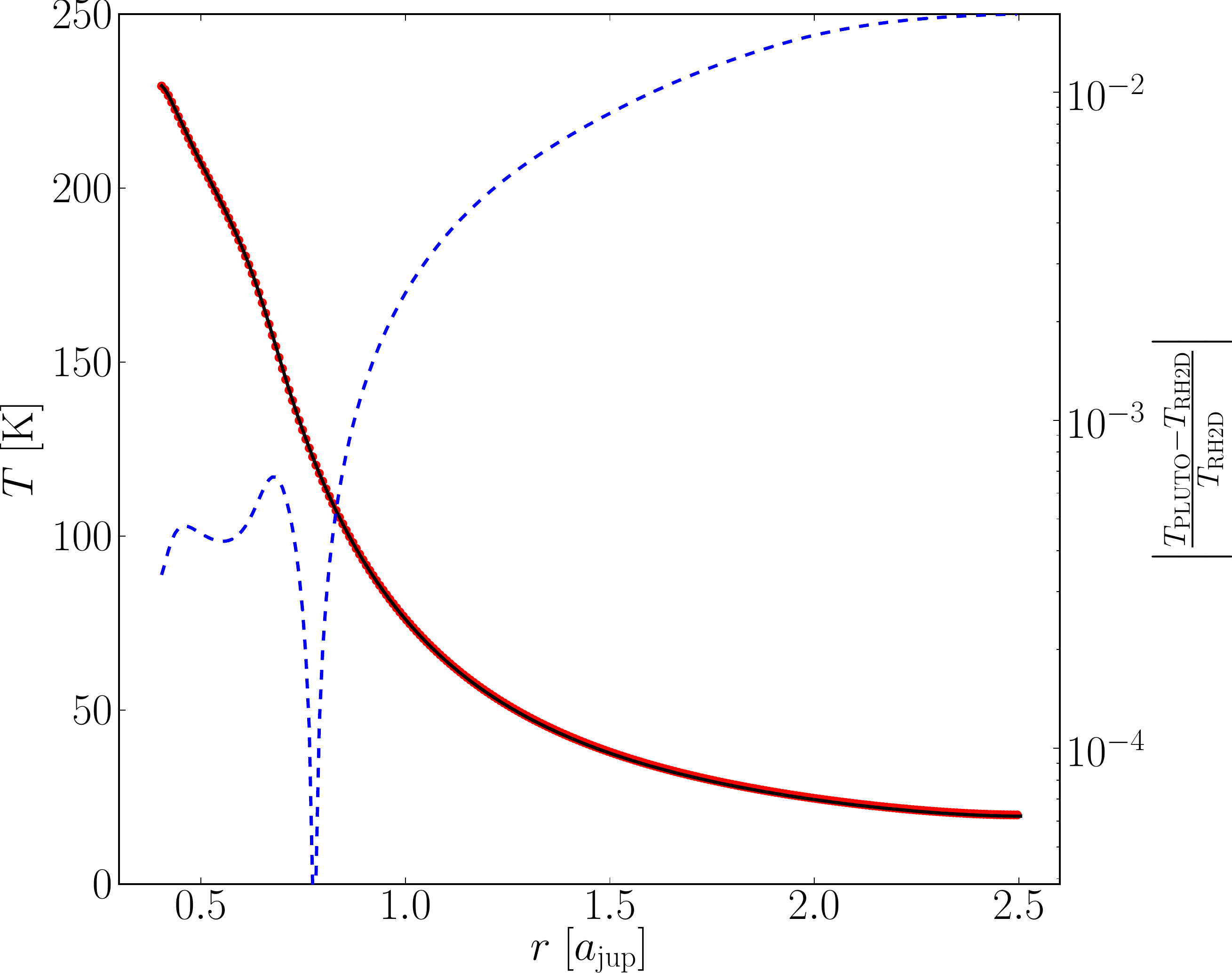}}
\caption{Radial mid-plane temperature profile in the simulations with \PLUTO (red dots) and 
with the code {\tt RH2D} (black line) after $t=\unit[10]{orbits}$ \subref{pic:accretion_disc_test:10orbits} and 
$t=\unit[100]{orbits}$ \subref{pic:accretion_disc_test:100orbits}, together with the absolute value of the relative error (blue dashed 
line) which belongs to the log axis on the right.}
\label{pic:accretion_disc_test}
\end{figure}

\subsubsection{The static case}\label{subsubsec:ad-static}
In this test case only the radiative equations are solved without the hydrodynamics.
In order to account for the viscous heating in this case, 
we add an additional dissipation contribution, $D$, to the right hand side of the internal energy equation in
\eqref{eqn:energy_coupling}. We consider standard viscous heating, and include only the main contribution due to the
approximately Keplerian shear flow. At the individual grid points the dissipation is then given by
\begin{equation}
   D_\ind = r_i^2 \rho_\ind \nu \left(\frac{\partial \Omega_\ind}{\partial r_i}\right)^2 \,,
\end{equation}
where $\nu$ is the constant viscosity and $\Omega_\ind$ the angular velocity at the individual grid points.
In summary we solve the same equations as in the case with irradiation, when we substitute $S_\ind$ with $D_\ind$.

\begin{figure}
\center
\subfigure[$\gamma = 5/3$]{\label{pic:full-disk-gam16}
  \includegraphics[width=0.4\iswidth]{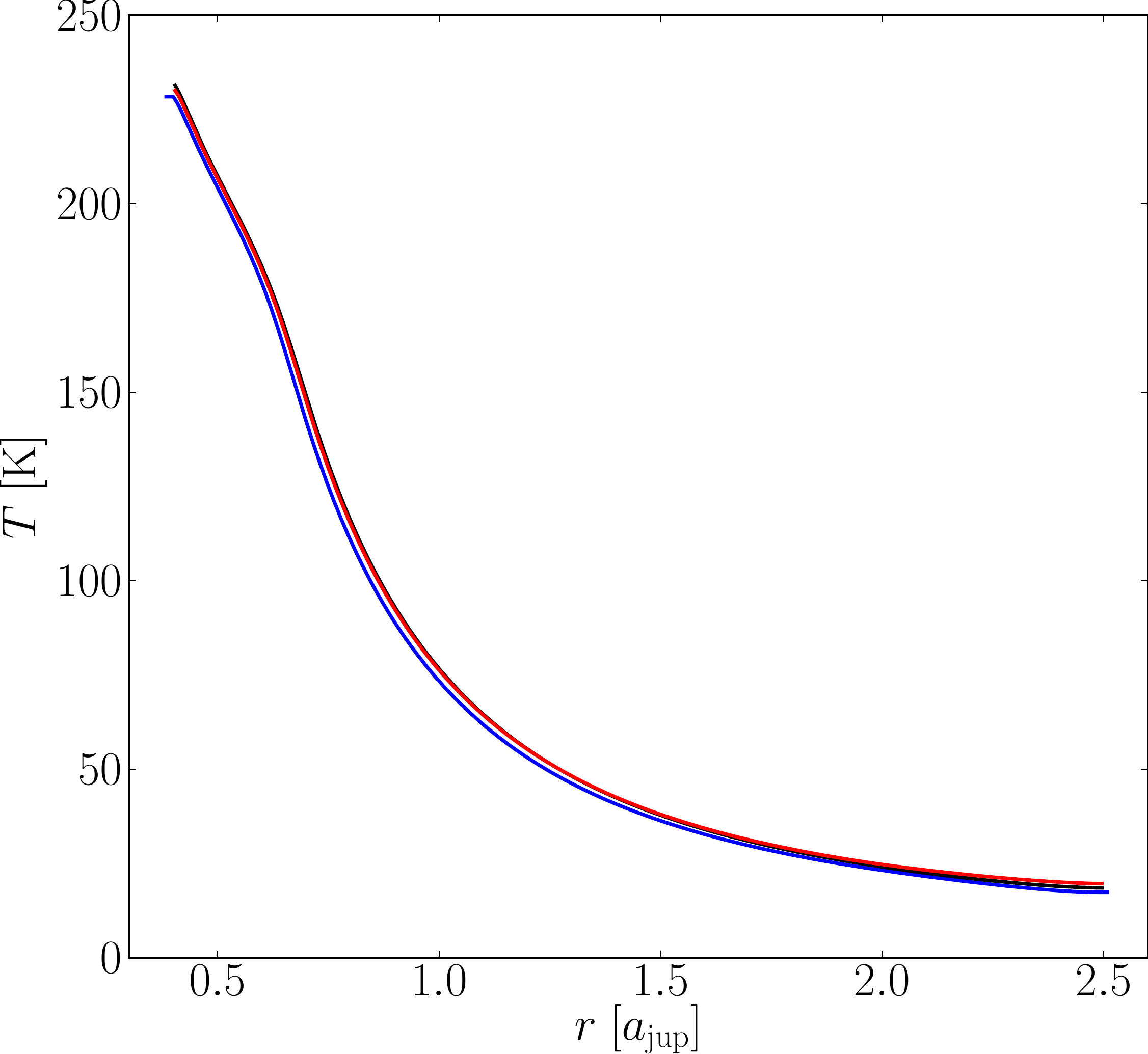}}
\subfigure[$\gamma = 1.1$]{\label{pic:full-disk-gam11}
 \includegraphics[width=0.4\iswidth]{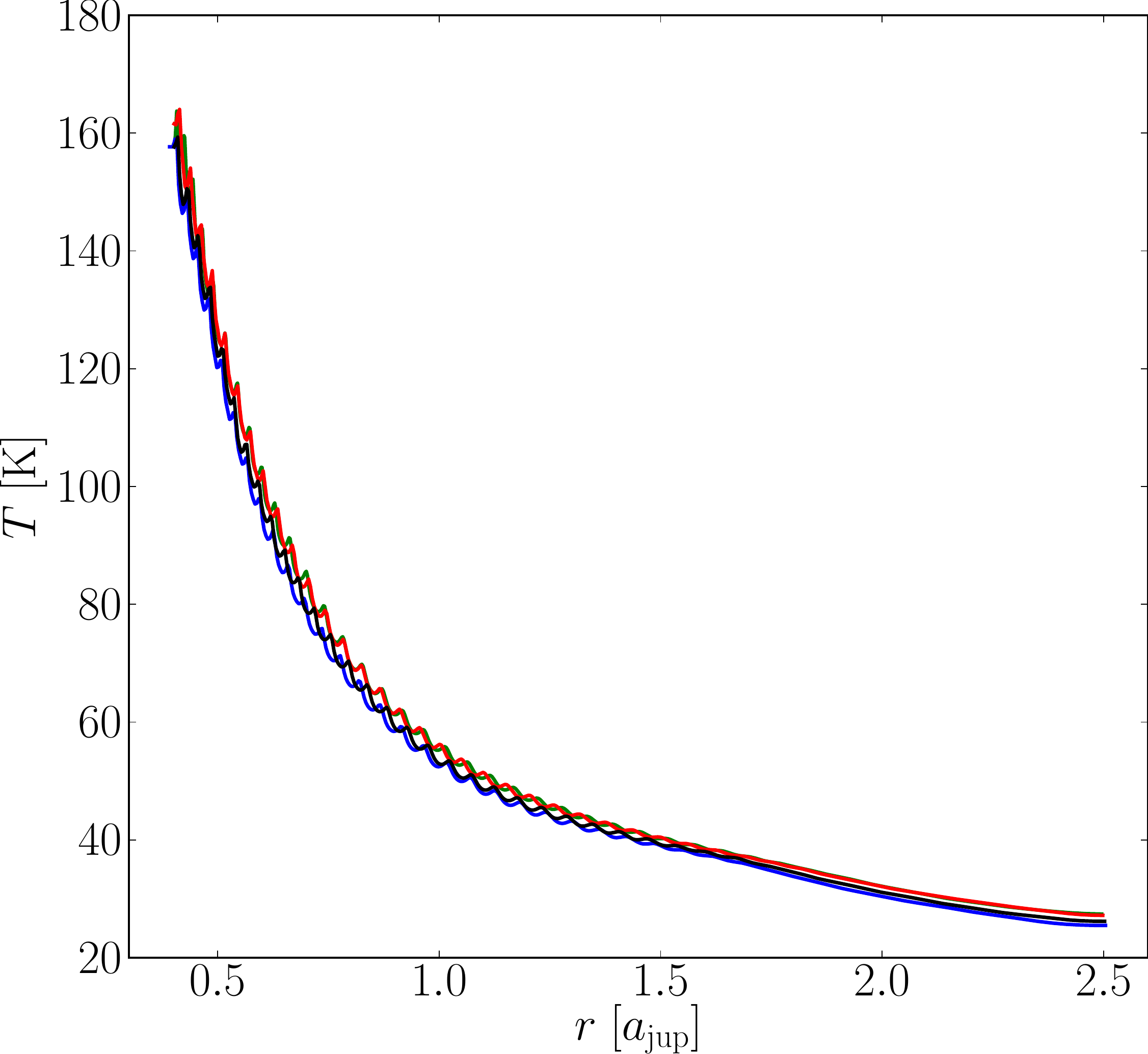}}
\caption{Radial mid-plane temperature profile in the simulation with \PLUTO (red line), {\tt RH2D} (black line)
and {\tt NIRVANA} (blue line) in the quasi-equilibrium state after 100 orbits in the
case with $\gamma = 5/3$ without convection \subref{pic:accretion_disc_test:10orbits} and in the strongly 
convective case with $\gamma =1.1$ \subref{pic:accretion_disc_test:100orbits}. Additionally we added the results
of a simulation performed with \PLUTO where we use a logarithmic grid in $r$-direction (green line).
}
\label{pic:full-disk-test}
\end{figure}

In the steady state, the time derivatives in the equations \eqref{eqn:energy_coupling}
vanish and the system is reduced to the following equation for the radiation energy density
\begin{equation}
  \nabla \cdot \, \left( \frac{c \lambda}{\kappa_\mathrm{R} \rho} \nabla E \right)  =  D.
  \label{eq:disc-equilib}
\end{equation}
In optically thick regions, $E=a_{\rm R} T^4$ and eq.~(\ref{eq:disc-equilib}) determines the temperature stratification
within the disc.

\begin{figure*}
\center
\includegraphics[width=0.9\ibwidth]{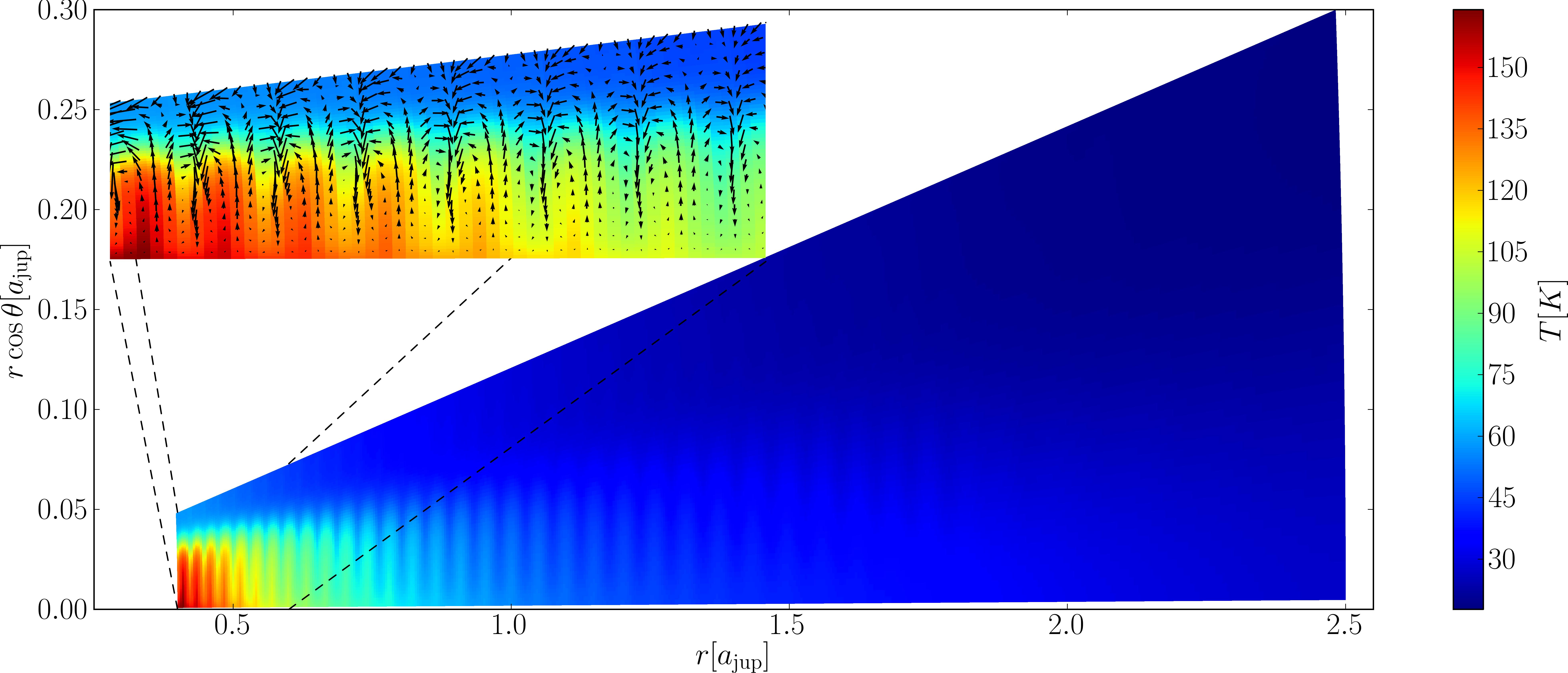}
\caption{Vertical slice of the disc temperature at $t=\unit[100]{orbits}$ in the dynamical case with $\gamma=1.1$ showing convection cells.
Also plotted in the inset is the enlarged region from $r=0.4\,a_{\text{jup}}$ to $0.6\,a_{\text{jup}}$ 
with the velocity field in the $r-\theta$ plane (black arrows).
}
\label{pic:full-disk-test:convection_cells}
\end{figure*}

The simulation starts at $t=\unit[0] {orbits}$ and is evolved until 
$t=\unit[100]{orbits}$ are reached, where one orbit corresponds to the Keplerian orbital period at the distance
of $a_\text{jup}$ which is given here by $\unit[3.732 \cdot 10^8]{s}$. The initial and overall time step was chosen as $\Delta t 
=\unit[10^{-3}]{orbits}=\unit[3.732 \cdot 10^{5}]{s}$. 
The results for the static case are shown in figure \ref{pic:accretion_disc_test} using here a value of $\gamma = 7/5$ for
the adiabatic index.
The plots show the radial temperature profile of 
the accretion disc in the mid-plane for the simulations after 10 orbits (top panel) and after 100 orbits (bottom panel).
We display results of two different simulations, one done with the code \PLUTO (red dots) using the described methods, 
and the second (black lines) run with the code {\tt RH2D} \citep{1989A&A...208...98K}.
The result from both codes are nearly identical.
Even after $\unit[100]{orbits}$ the absolute value of the relative error is always less than $2\%$.
The test shows that the time-scale of the radiative evolution, as well as the equilibrium state is captured correctly. 
We note that the code {\tt RH2D} uses the one-temperature approach of radiation transport in this case.

\subsubsection{The dynamical case} \label{sec:accretion_disc_dynamic_case}

The final equilibrium of the described static case does not depend on the magnitude of $\gamma$, because the
viscous heating is independent of it, see eq.~(\ref{eq:disc-equilib}). 
The situation is different, however, for the dynamical cases, where the
hydrodynamical evolution of the flow is taken into account. Since the time scale of the radiative transport depends on
$\gamma$ (through eq.~\ref{eq:cv}),
one might expect the possibility of convective instability, see for example the recent work by \cite{2013A&A...550A..52B}.
This is indeed the case for small enough values of $\gamma$. In order to demonstrate the correctness of our implementation also
for the full dynamical problem, we modelled two discs, one with $\gamma = 5/3$ which clearly shows no convection, and the other
with $\gamma = 1.1$ which shows strong convection. The initial setup was identical to that described before, but now we solve the 
equations of viscous hydrodynamics with radiation transport, but without irradiation and explicit dissipation.
Please note that for viscous flows the energy generation due to viscous dissipation is automatically included in the total
energy equation.
The equations \eqref{eqn:continuity_equation} to \eqref{eqn:gas_energy_equation_coupled} are solved by \PLUTO, and
the system of equations \eqref{eqn:energy_coupling} are solved as described in 
section \ref{sec:solving_the_radiation_part}. Since this setup is very dynamical
and requires a more complex interplay of hydrodynamics and radiative transport, we use an additional third code, {\tt NIRVANA},
for comparison. The {\tt NIRVANA} code has been used in \citet{2009A&A...506..971K} and \citet{2013A&A...550A..52B}
on very similar setups.
The results of the two cases are shown in Fig.~\ref{pic:full-disk-test}. In the top panel (a)
we display the result for the $\gamma = 5/3$ case which is not convective. Here, the agreement between the
codes is excellent with the maximum deviation in the percentage range.
In the lower panel (b) we display the results for the $\gamma = 1.1$ case. Here the radiative transport time-scale is
enhanced which leads to a strongly convective situation, which can be seen in the raggedness of the curves.
In this simulation we doubled the spatial resolution, compared with the $\gamma=5/3$ case,
such that the convection cells are reasonably well resolved, see figure \ref{pic:full-disk-test:convection_cells}.
The agreement between the three different codes is very good, despite of the very different solution methods
for the hydrodynamics equations: {\tt PLUTO} uses the total energy equation with a Riemann-solver while {\tt RH2D} and {\tt NIRVANA}
use a second-order upwind scheme and the thermal energy equation.
Additionally, the latter two codes use the full dissipation function and the one-temperature approach.

\subsubsection{Parallel scaling} \label{subsubsec:parallel_scaling}

\begin{figure}
\center
\includegraphics[width=0.41\iswidth]{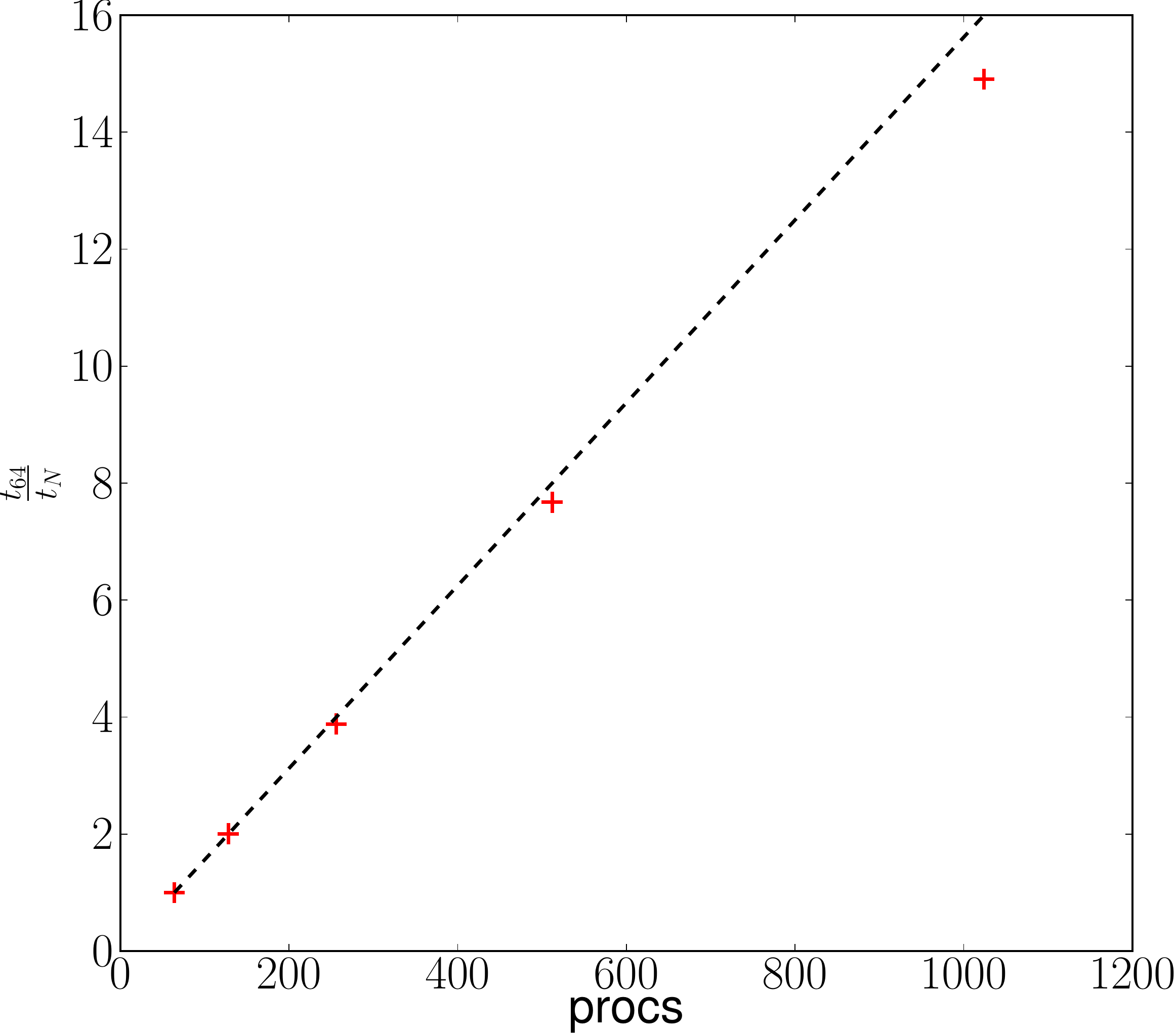}
\caption{Parallel scaling benchmark results for the static accretion disc test case. We plot here the 
number of processor cores against $\frac{t_{64}}{t_\mathrm{N}}$ where $t_{N}$ is the runtime used on 
$N$ processors accordingly for $t_{64}$. The used run-times with full hydrodynamics and radiation transport for 
$64$, $128$, $256$, $512$ and $1024$ cpu cores (red crosses) are shown together
with the ideal case (black dashed line).}
\label{pic:parallel_scaling:accretion_disc}
\end{figure}

In order to test the parallel scaling of our new implementation, we used the same setup as in section \ref{sec:accretion_disc_dynamic_case} and
increased the number of grid cells to $1024 \times 64 \times 256$. 
The computations were only run until $t=\unit[5]{orbits}$, and we used the solver PETSc. So we were able to run the test on 
$64$ up to $1024$ processor cores within a reasonable time.
The simulations were run on clusters of the BWGrid which are equipped with Intel Xeon E5440 cpus and have a low latency InfiniBand network.
In figure \ref{pic:parallel_scaling:accretion_disc} we show
the results of the simulations performed with full hydrodynamics and radiation transport.
The run-time increases nearly by a factor of two when doubling the number of cores.
With this setup, solving the hydrodynamics equations needs between $40\%$ and $50\%$ of the computation time and the radiation 
transport the remaining $60\%$ to $50\%$, however, these numbers are strongly problem-dependent.
Therefore even up to 1024 cores, we see good agreement with ideal scaling. According to 
Amdahl's law the full code, including the original part of \PLUTO and our 
implementation of the radiation transport, is well parallelised.

\section{Summary and conclusions}
We described the implementation of a new radiation module to the {\tt PLUTO} code. The module solves for the flux-limited 
diffusion approximation in the two-temperature approach. For discretisation the finite volume method is used, and
the resulting difference equations couple the updates of the temperature and radiation energy density.
Due to possibly severe time step limitations, the set of equations is solved implicitly. For treating the non-linearity of the temperature
in the matter-radiation coupling term, we utilize the method of \citet{2011A&A...529A..35C}.

The accuracy of the implementation has been verified using different physical and numerical setups.  
The first set of tests deals with purely radiative problems that include the purely diffusive evolution towards 
an equilibrium, and special setups to test the coupling terms between radiative and thermal energy.
A newly developed setup checks for the correct inclusion of the irradiation from a central source in a spherical coordinate system.

In the second test suite we study the full simultaneous evolution of hydrodynamics and radiation.
First, sub- and super-critical radiative shock simulations are performed and their outcomes agree
very well with published results of identical setups. Finally, we study the onset of convection in internally heated
viscous discs, and find very good agreement between 3 different, independent hydrodynamical codes.
This last test also allowed us to test the correct implementation in a spherical coordinate system and a non-equidistant
logarithmic grid. Our numerical performance tests indicate excellent parallel scaling, up to at least 1024 processors.

The current version of the radiation module comes with routines for the Rosseland mean opacity from 
\cite{1985prpl.conf..981L} and \cite{1994ApJ...427..987B}. 
Additionally it is possible to use the Rosseland and Planck mean opacities from \cite{2003A&A...410..611S}.
 
The described radiation module can be easily used within the \PLUTO-environment. It can be found on the webpage 
\footnote{\url{http://www.tat.physik.uni-tuebingen.de/~pluto/pluto_radiation/}} as
a patch for the version 4.0 of \PLUTO.

\begin{acknowledgements}
We gratefully thank the bwGRiD project\footnote{bwGRiD (http://www.bw-grid.de), member of the 
German D-Grid initiative, funded by the Ministry for Education and Research (Bundesministerium 
f\"ur Bildung und Forschung) and the Ministry for Science, Research and Arts Baden-W\"urttemberg 
(Ministerium f\"ur Wissenschaft, Forschung und Kunst Baden-W\"urttemberg).} for the computational 
resources.
We gratefully acknowledges support through the German Research Foundation (DFG) through grant KL 650/11
within the Collaborative Research Group FOR 759: {\it The formation of Planets: The Critical First Growth Phase}. 
We thank Rolf Kuiper for many stimulating discussions, either physical or technical.
\end{acknowledgements}
\bibliography{references}{}
\bibliographystyle{aa}
\end{document}